\pdfoutput=1

\documentclass[showpacs,prl,twocolumn,aps]{revtex4}

\usepackage{bm}
\usepackage[normalem]{ulem}
\usepackage{graphicx}
\usepackage{amsmath}
\usepackage{amssymb}
\usepackage{latexsym}
\usepackage{amsfonts}
\usepackage{epsfig}
\usepackage{pstricks}
\usepackage{psfrag}
\usepackage{exscale}
\usepackage{times}
\usepackage{color}

\def\DJo{$\;$\kern-.4em \hbox{D\kern-.8em\raise.15ex\hbox{--}\kern.35em okovi\'c}}
\newcommand{\ket}[1]{\left|#1\right>}
\newcommand{\bra}[1]{\left<#1\right|}   

\newcommand{\nn}{\nonumber\\}

\newcommand{\bea}{\begin{eqnarray}}
\newcommand{\ea}{\end{eqnarray}}
\newcommand{\eea}{\end{eqnarray}}

\begin{document}

\title{Avalanche of entanglement and correlations at quantum phase transitions} 


\author{Konstantin V. Krutitsky}
%
\author{Andreas Osterloh}
%
\author{Ralf Sch\"utzhold}
\affiliation{Institut f\"ur Theoretische Physik, 
         Universit\"at Duisburg-Essen, D-47048 Duisburg, Germany.}


\begin{abstract}
We study the ground-state entanglement in the quantum Ising model 
with nearest neighbor ferromagnetic coupling $J$
and find a sequential increase of entanglement depth with growing $J$.
This entanglement avalanche starts with two-point entanglement, 
as measured by the concurrence, 
and continues via the three-tangle and four-tangle, until finally, 
deep in the ferromagnetic phase for $J=\infty$, 
arriving at pure $\ell$-partite (GHZ type) entanglement of 
all $\ell$ spins.
Comparison with the two, three, and four-point correlations reveals a similar 
sequence and shows strong ties to the above entanglement measures 
for small $J$. 
However, we also find a partial inversion of the hierarchy, 
where the four-point correlation exceeds the three- and two-point correlations, 
well before the critical point is reached. 
Qualitatively similar behavior is also found for the Bose-Hubbard model,
suggesting that this is a general feature of a quantum phase transition.
This should have far reaching consequences for approximations 
starting from a mean-field limit.
\end{abstract}
\pacs{03.67.Mn, 
05.30.Jp, 
05.30.Rt, 
75.10.Pq 
}

\maketitle


\paragraph{Introduction}
%
Entanglement is one of the main reasons for the complexity of quantum many-body
systems in that, the more entangled a quantum system is the more 
complex its description becomes. 
As an example, let us consider ground states of lattice Hamiltonians.
If we have no entanglement between the lattice sites $i$, the quantum state 
is fully separable $\ket{\Psi}=\bigotimes_i\ket{\psi_i}$ and thus quite 
simple. 
As a result, it is possible to employ a mean-field description where 
observables $\hat A_i$ and $\hat B_j$ at different lattice sites 
$i$ and $j$ are uncorrelated
$\langle\hat A_i\hat B_j\rangle=
\langle\hat A_i\rangle\langle\hat B_j\rangle$.
However, except for very few special cases 
(mean-field limit or Kurmann-Thomas-M\"uller point~\cite{Kurmann,Illuminati-FF09}),  
such a description is not exact. 
It is possible to improve this mean-field ansatz by adding some amount of 
entanglement. 
This can be achieved either directly~\cite{SchuetzOst15} or with matrix product 
states~\cite{MPS,VerstraeteConti10} or tree-tensor 
networks~\cite{PizornVerstraete13,Tagliacozzo09,ShiDuanVidal06}.
But these descriptions do only work reliably if the entanglement is bounded 
in a suitable way as highlightd in Refs.~\cite{dmrg1,footnote1}.

On the other hand, many interesting phenomena in condensed matter are 
associated with and occur at or close to quantum critical points, where 
typically the entanglement becomes very large.
As one of the simplest yet prototypical examples~\cite{Sachdev99}, 
let us consider the one-dimensional Ising model in a transverse field
\begin{eqnarray}
\label{Ising}
\hat H
=
-J\sum_{i=1}^\ell\hat\sigma_i^z\hat\sigma_{i+1}^z 
-\sum_{i=1}^\ell\hat\sigma_i^x
\,,
\end{eqnarray}
where $\hat\sigma_i^{x,y,z}$ denote the spin-$1/2$ Pauli matrices acting on the 
lattice site $i$ and periodic boundary conditions 
$\sigma^z_{\ell+1}=\sigma^z_{1}$ 
are imposed. 
This model displays a ${\mathbb Z}_2$ symmetry corresponding to the 
simultaneous flip of all $\ell$ spins. 
In this regard, for $\ell\to\infty$, we have a symmetry-breaking 
second-order quantum phase transition from the paramagnetic phase 
at $|J|<1$ to ferromagnetism at $J>1$~\cite{Sachdev99}.
For $J=0$, we have the separable paramagnetic state $\ket{\to\to\to\dots}$ 
without entanglement while for $J\to\infty$, the ground state corresponds 
to the ferromagnetic state $(\ket{\uparrow\uparrow\uparrow\dots}+
\ket{\downarrow\downarrow\downarrow\dots})/\sqrt{2}$
with GHZ-type multi-partite entanglement between all $\ell$ spins.
Here, $\ket{\to}$ is the eigenstate of $\sigma^x$ 
and $\ket{\uparrow}$ that of $\sigma^z$ to the eigenvalue $+1$.
At the critical point $J_{\rm crit}=1$, the entanglement entropy between the 
left and the right half of the Ising chain diverges as 
$\ln\ell$~\cite{GVidal3}.

\begin{figure}
\centering

\includegraphics*[width=.9\linewidth]{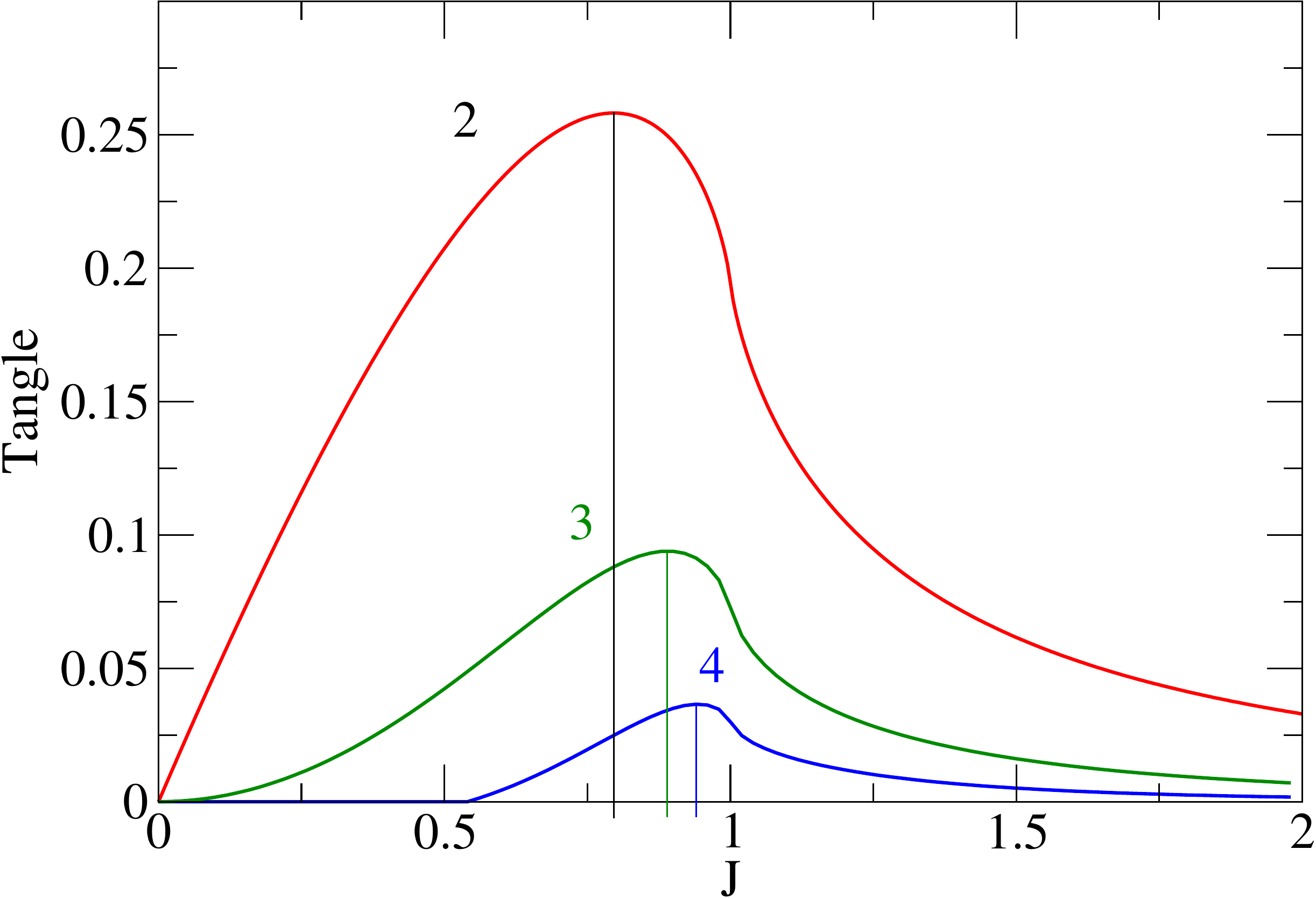}

\caption{(color online) The entanglement between two, three, and four neighboring spins 
measured by the concurrence $C_2$ (black), the three-tangle $\sqrt{\tau_3}$ 
(green), and the four-tangle $\tau_4$ (blue) as a function of $J$. 
For the latter two, the approximation~\eqref{approximation} was used. 
The concurrence starts linearly for small $J$ and the three-tangle 
$\sqrt{\tau_3}$ quadratic, while the four-tangle vanishes until
$J_0\approx 0.55$. 
They assume their maximum values at $J_2^{\rm max}\approx 0.796$, 
$J_3^{\rm max}\approx 0.890$, and $J_4^{\rm max}\approx 0.94$, respectively -- 
which shows the sequential increase of entanglement depth 
(avalanche of entanglement).}

\label{tangles}
\end{figure}

However, this large amount of entanglement cannot be explained by 
entanglement of pairs alone~\cite{amico04}, as measured by the concurrence.
Together with the entanglement monogamy relation~\cite{Coffman00,Osborne06}
(see also~\cite{AdessoRegula14,AdessoRegula14b,AdessoRegula16,AdessoOsterlohRegula16})
this strongly suggests the emergence of multipartite entanglement~\cite{Hofmann14,OstUpperBound16} 
(triples and quadruples etc.), 
which will be studied in the following section, see also Fig.~\ref{tangles}.
The relation to multipartite quantum correlations will be analyzed later on,  
also with reference to another prototypical model of quantum 
phase transitions, the Bose-Hubbard model.

\paragraph{Entanglement}
%
In order to study the multi-partite entanglement during the quantum phase 
transition of the Ising model, we employ its exact solution via Jordan-Wigner 
and Bogoliubov transformation to a free fermionic model~\cite{PFEUTY,LIEB} 
which allows us to obtain the reduced density matrices of two 
$\hat\rho_2=\hat\rho_{ij}$, three $\hat\rho_3=\hat\rho_{ijk}$, and four 
$\hat\rho_4=\hat\rho_{ijkl}$ neighboring spins~\cite{PFEUTY,Hofmann14}.
After diagonalizing these matrices, we find that they all possess two 
dominant eigenvalues $p_1$ and $p_2$ while the sum of the remaining 
sub-dominant eigenvalues stays below $2.5\%$ 
(for details, see the supplement~\cite{Supplement}).
Thus, we approximate the two-point reduced density operators as
\begin{eqnarray}
\label{approximation}
\hat\rho_{ij}
\approx
p_1\ket{\psi_{ij}^1}\bra{\psi_{ij}^1}
+
(1-p_1)\ket{\psi_{ij}^2}\bra{\psi_{ij}^2} 
\;,
\end{eqnarray}
and analogously for $\hat\rho_{ijk}$ and $\hat\rho_{ijkl}$.
Actually, the accuracy of this approximation should be even better 
than $2.5\%$:
while the multi-partite entanglement of the first $\ket{\psi_{\dots}^1}$ and 
the second $\ket{\psi_{\dots}^2}$ eigenvectors can interfere destructively
with each other, we checked that this is not the case for the third. 
The third eigenvector $\ket{\psi_{\dots}^3}$ has a different structure:
For three and four spins, the state(s) of the central spin(s) are fixed to 
$\ket{\to}$ while the two boundary spins form a Bell state -- i.e., 
$\ket{\psi_{\dots}^3}$ contains bi-partite entanglement only, which here does not 
interfere with the multi-partite entanglement of $\ket{\psi_{\dots}^1}$ and
$\ket{\psi_{\dots}^2}$.  
As a result, we expect that the accuracy of this approximation is around 
$0.5\%$ or even better. 
For two spins, we checked this approximation by 
comparing the exact concurrence with that derived from~\eqref{approximation}
and found that they are virtually indistinguishable 
(see the supplement~\cite{Supplement}).

The approximation~\eqref{approximation} as motivated by the dominance of 
the two largest eigenvalues is a great simplification, because we obtain 
rank-two density matrices, for which the three-tangle $\tau_3$ and the 
four-tangle(s) $\tau_4$ can be calculated exactly~\cite{LOSU,KENNLINIE} 
for this model.
Note that an exact extension to arbitrary
mixed states by the convex roof is not known so far for the three-tangle
since the homogeneity degree of the polynomial measure is larger than two.  

In analogy to the three-tangle $\tau_3$, we call those polynomial 
$SL$-invariants 
that are zero for arbitrary product states~\cite{OS04} four-tangle 
and use the notation $\tau_4^{(i)}$, $i=1,2,3$, 
for those powers that scale linearly in the density matrix.
All three of them essentially lead to the same output, and therefore 
$\tau_4$ will represent
the four-partite entanglement content of the model.
This four-partite entanglement is hence of GHZ-type because only the GHZ 
entanglement is measured 
by all three measures in the same way~\cite{OS04,OS05,DoOs08}.
Altogether, these quantities $\tau_3$ and $\tau_4$ measure the 
tri-partite entanglement 
of $\hat\rho_{3}$ and the quadri-partite entanglement of 
$\hat\rho_{4}$, and are shown in Fig.~\ref{tangles}. 
We plotted $\tau_3^{1/2}$ because this quantity also yields a homogeneous 
functions of degree one in the density matrix $\hat\rho_3$
and thus all the tangles shown have properties similar to probabilities.
For completeness, we also included the pairwise entanglement 
of nearest neighbors, as measured by the concurrence $C_2$.

As is well-known, the concurrence first grows as a function of $J$
until it reaches a maximum at $J\approx 0.796$ and later decreases again
(with an infinite slope at the critical point~\cite{OstNat}).
The three-tangle $\sqrt{\tau_3}$ starts to grow much slower
at small $J$ and reaches its maximum later than the concurrence at 
$J\approx 0.89$.
The four-tangle(s) $\tau_4$ are even zero until $J\approx 0.55$ and reach 
their maximum yet a bit later at $J\approx 0.94$.
Even though having no results for more spins, we conjecture that this 
sequence or avalanche of entanglement continues until finally, 
deep in the ferromagnetic phase, we get pure $\ell$-partite entanglement 
of all spins~\cite{footnote2}.

\begin{figure}[t]

\centering

\includegraphics*[width=.9\linewidth]{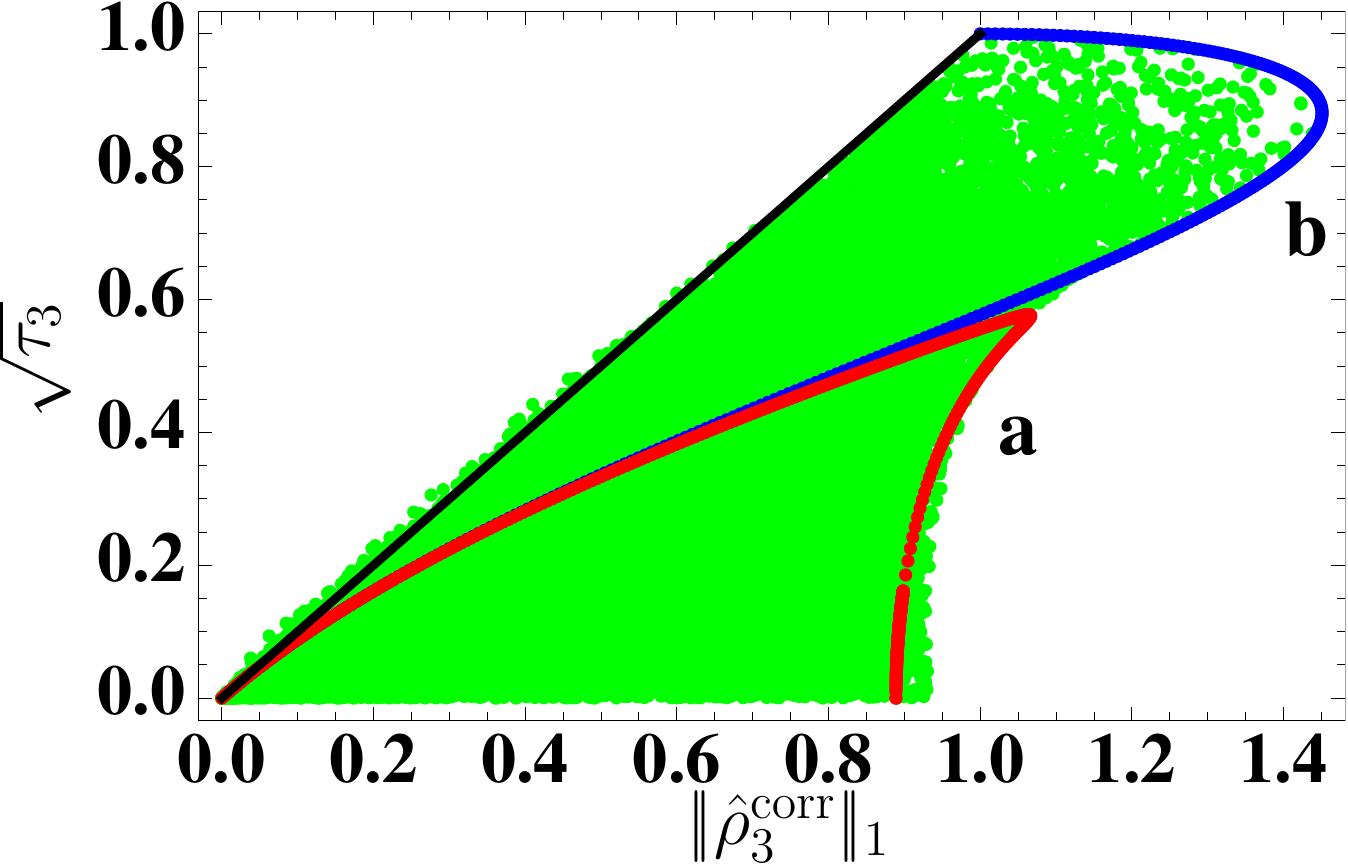}

\caption{
(color online)
Plot of the three-tangle $\sqrt{\tau_3}$ versus the three-point 
correlation bound $||\rho_3^{\rm corr}||_1$ for a random selection of 40.000
pure Acin states~\cite{Acin01}.
Each dot in the figure corresponds to a single state. 
The thin black line corresponds to 
$\sqrt{\tau_3}=|\!|\hat\rho_3^{\rm corr}|\!|_1$.
The lower red (a) and upper blue curve (b) are obtained for GHZ states 
that are made of two
($\alpha|\!\!\uparrow\uparrow\uparrow\rangle+\beta|\!\!\downarrow\downarrow\downarrow\rangle$)
and four
($\alpha|\!\!\uparrow\uparrow\uparrow\rangle+\beta|{\rm W}\rangle$)
product basis elements, respectively, where
$|W\rangle=({|\!\!\downarrow\uparrow\uparrow\rangle}+{|\!\!\uparrow\downarrow\uparrow\rangle}+{|\!\!\uparrow\uparrow\downarrow\rangle})/\sqrt{3}$.
Although $||\rho_3^{\rm corr}||_1$ is not always larger than $\sqrt{\tau_3}$
(as it is the case for two spins), it is however satisfied approximately.
}
\label{tau3-of-rho3}
\end{figure}

\paragraph{Entanglement versus correlations}
%
As already mentioned in the Introduction, a pure state without any 
entanglement is fully separable $\ket{\Psi}=\bigotimes_i\ket{\psi_i}$ 
and thus quite simple. 
For example, observables $\hat A_i$ and $\hat B_j$ at different lattice 
sites $i$ and $j$ are uncorrelated
$\langle\hat A_i\hat B_j\rangle=
\langle\hat A_i\rangle\langle\hat B_j\rangle$.
In the following, we shall study the relation between entanglement and 
the resulting correlations in more detail.
To this end, we start with the reduced density matrices for one 
$\hat\rho_1=\hat\rho_{i}$, two $\hat\rho_2=\hat\rho_{ij}$, 
three $\hat\rho_3=\hat\rho_{ijk}$, and four $\hat\rho_4=\hat\rho_{ijkl}$ 
spins and split up the correlated parts via 
$\hat\rho_{ij}^{\rm corr}=\hat\rho_{ij}-\hat\rho_{i}\hat\rho_{j}$, and 
analogously for more spins (see supplement~\cite{Supplement}).
The correlation between the two observables 
$\langle\hat A_i\hat B_j\rangle^{\rm corr}=
\langle\hat A_i\hat B_j\rangle-
\langle\hat A_i\rangle\langle\hat B_j\rangle$
can be written as 
$\langle\hat A_i\hat B_j\rangle^{\rm corr}=
{\rm Tr}\{\hat A_i\hat B_j\hat\rho_{ij}^{\rm corr}\}$
and similarly for three or more sites.

Since correlations such as $\langle\hat A_i\hat B_j\rangle^{\rm corr}$ and 
$\langle\hat A_i\hat B_j\hat C_k\rangle^{\rm corr}$ obviously depend on 
the observables $\hat A_i$, $\hat B_j$, and $\hat C_k$, it is convenient 
to derive an estimate directly from the correlated density matrices such as 
$\hat\rho_{ij}^{\rm corr}$. 
For observables whose norm is bounded by unity 
($|\hat A_i|\leq1$, $|\hat B_j|\leq1$ etc.) such as the Pauli spin 
matrices, we obtain
\bea
\label{Schatten}
\langle\hat A_i\hat B_j\rangle^{\rm corr}
&=&
{\rm Tr}\left\{\hat A_i\hat B_j\hat\rho_{ij}^{\rm corr}\right\}
=
\sum_I\lambda_I\bra{\chi_{ij}^I}\hat A_i\hat B_j\ket{\chi_{ij}^I}
\nn
&\leq&
\sum_I|\lambda_I|=||\hat\rho_{ij}^{\rm corr}||_1
\,,
\ea
where we have inserted the diagonalization of $\hat\rho_{ij}^{\rm corr}$
with eigenvalues $\lambda_I$ and eigenvectors $\ket{\chi_{ij}^I}$.
We find that the Schatten one-norm $||\hat\rho_{ij}^{\rm corr}||_1$ of the 
correlated density matrix $\hat\rho_{ij}^{\rm corr}$ yields an upper estimate 
for the correlations $\langle\hat A_i\hat B_j\rangle^{\rm corr}$
of all observables whose norm is bounded by unity.  
Hence, we shall focus on this quantity in the following. 
Obviously, the same argument can be applied to three or more sites
in complete analogy. 

For two spins, it is well-known that the largest correlation function 
for pure states coincides with the concurrence~\cite{VerstraetePC04}. 
For mixed states, this becomes an upper bound, i.e., the maximum correlation 
is larger or equal to the concurrence 
$||\hat\rho_{ij}^{\rm corr}||_1\geq C_2$.
Unfortunately, for three or more spins, such a rigorous bound is not known.

Thus, let us consider a system of three spins.
A pure state of the system can be represented as a superposition of five local 
bases product states~\cite{Acin01}. 
We generate these states randomly and calculate the three-tangle 
$\tau_3$ and $||\hat\rho^{\rm corr}_{ijk}||_1$.
The results are plotted in Fig.~\ref{tau3-of-rho3}.
Again, we show $\sqrt{\tau_3}$ because this quantity is a homogeneous 
function of degree one in the density operator
$\hat\rho_3$ and thus has properties similar to a probability.
We find that $\sqrt{\tau_3}$ is a very good approximation for a lower 
bound to $||\hat\rho^{\rm corr}_{ijk}||_1$. 
This is very similar to the two-spin case above, but one should be aware 
that $\sqrt{\tau_3}$ can exceed $||\hat\rho^{\rm corr}_{ijk}||_1$ a little bit 
-- there are points in Fig.~\ref{tau3-of-rho3} which lie slightly on the 
left of the black diagonal line 
(indicating the points where $||\hat\rho^{\rm corr}_{ijk}||_1=\sqrt{\tau_3}$).

Similar calculations for four spins indicate that 
$||\hat\rho^{\rm corr}_{ijkl}||_1$ is also an approximate
upper bound for the four-tangle $\tau_4$. 
However, since the available phase space for four spins is much larger,
the statistics is rather poor 
(see more details in the supplement~\cite{Supplement}). 

\begin{figure}
\centering


\includegraphics*[width=.9\linewidth]{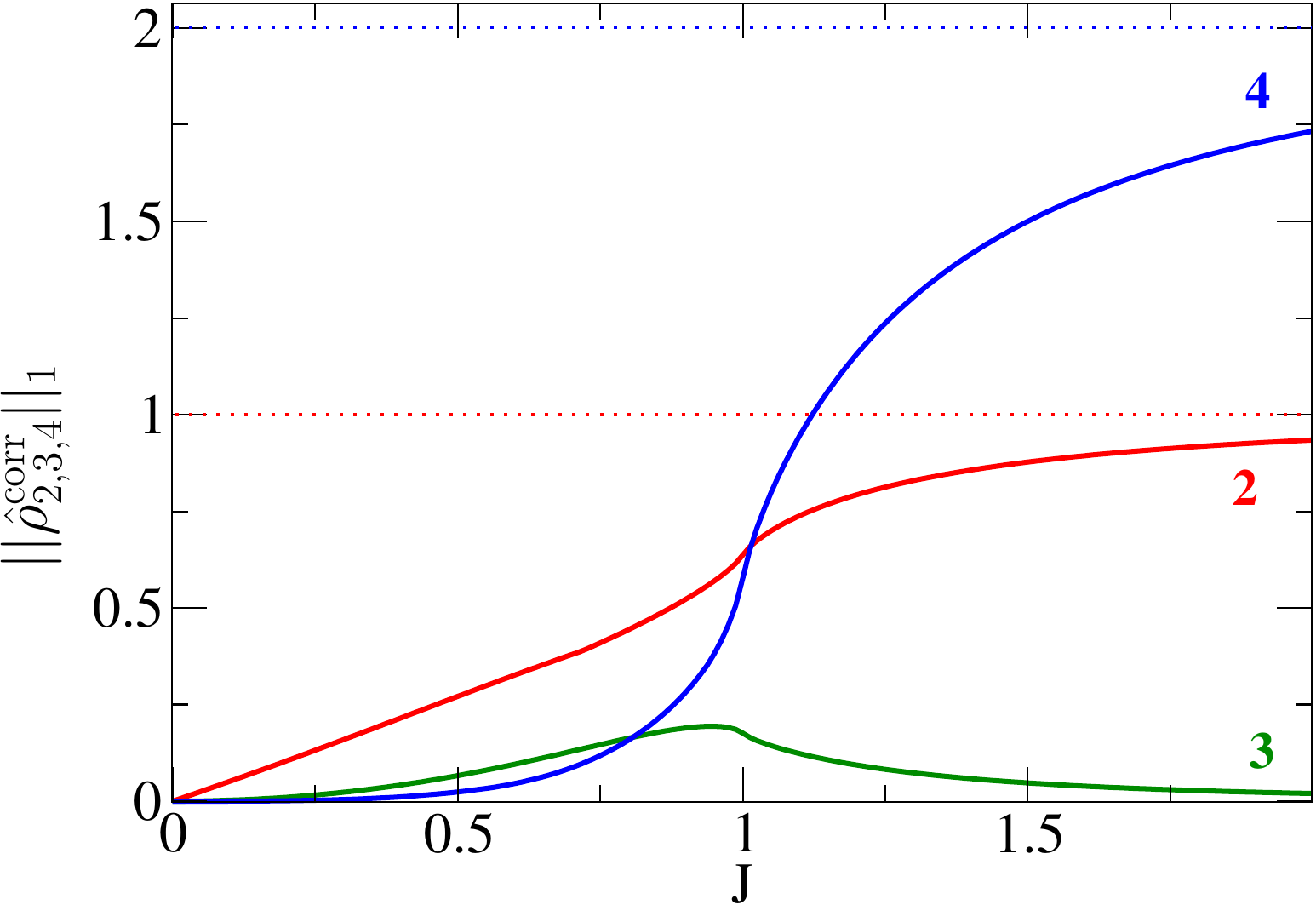}

\caption{(color online) Norms of correlated reduced density operators
for two $\hat\rho_{2}^{\rm corr}$ (red),
three $\hat\rho_{3}^{\rm corr}$ (green),
and four $\hat\rho_{4}^{\rm corr}$ (blue) 
neighboring spins in the transverse Ising model.
At $J\approx0.8$, i.e., well before the critical point, the 4-point 
correlations exceed the 3-point correlations.
The 2-point correlations dominate both until the critical point 
is reached, afterwards the 4-point correlations prevail. 
The horizontal dashed lines represent the asymptotic values for $J\to\infty$.}
\label{norm1.2to4}
\end{figure}

In summary, while the two-point correlation $\hat\rho_{ij}^{\rm corr}$
and the pairwise entanglement $C_2$ are related via the {\em exact} 
bound $||\hat\rho_{ij}^{\rm corr}||_1\geq C_2$, we find analogous approximate
relations between the three- and four-point correlations 
$\hat\rho_{ijk}^{\rm corr}$ and $\hat\rho_{ijkl}^{\rm corr}$ on the one hand 
and the corresponding entanglement measures $\sqrt{\tau_3}$
and $\tau_4$ on the other hand. 

\paragraph{Correlations for the Ising model}
%
Motivated by the above findings, let us study the Schatten one-norms of 
the correlated density matrices for two, three, and four neighboring spins. 
Note that we used the exact results for the reduced density matrices 
(obtained by Jordan-Wigner and Bogoliubov transformation) without the 
approximation~\eqref{approximation}.
The results are plotted in Fig.~\ref{norm1.2to4}.
As expected from stationary perturbation theory (see supplement), the 
two-point correlation $||\hat\rho_{2}^{\rm corr}||_1=||\hat\rho_{ij}^{\rm corr}||_1$
behaves linearly in $J$ for small $J$, 
while the three-point correlation 
$||\hat\rho_{3}^{\rm corr}||_1=||\hat\rho_{ijk}^{\rm corr}||_1$ 
and the four-point correlation 
$||\hat\rho_{4}^{\rm corr}||_1=||\hat\rho_{ijkl}^{\rm corr}||_1$ 
scale with the second and third power of $J$, respectively. 

Thus, for small $J$, the correlations obey the same hierarchy 
$||\hat\rho_{2}^{\rm corr}||_1\gg||\hat\rho_{3}^{\rm corr}||_1
\gg||\hat\rho_{4}^{\rm corr}||_1$ as the entanglement measures in
Fig.~\ref{tangles}, except that $||\hat\rho_{4}^{\rm corr}||_1$ 
does not vanish for finite $J$ in contrast to $\tau_4$. 
However, at $J\approx0.8$, i.e., well before the critical point, 
this hierarchy is violated as the four-point correlation 
$||\hat\rho_{4}^{\rm corr}||_1$ exceeds the three-point correlation 
$||\hat\rho_{3}^{\rm corr}||_1$.
The two-point correlation $||\hat\rho_{2}^{\rm corr}||_1$ is still dominant 
in this region -- this only changes near the critical point. 
This inversion of the hierarchy, i.e., the dominance of 
$||\hat\rho_{4}^{\rm corr}||_1$ over $||\hat\rho_{3}^{\rm corr}||_1$ 
in a region within the symmetric paramagnetic phase, 
should be relevant for approximation schemes 
which truncate the hierarchy of correlations at some order (see below). 

\paragraph{Bose-Hubbard model}
%
One might suspect that this inversion of the hierarchy is a rather specific 
result due to the integrability of the model under consideration or may be 
induced by the fact that deep in the ferromagnetic (broken-symmetry) phase, 
the three-point correlation vanishes whereas the four-point (and two-point) 
correlators approach constant non-zero values. 
(Note that an inversion of the two-point and four-point correlations
just happens at the critical point.) 
In order to investigate whether the inversion of the hierarchy is a general 
phenomenon or indeed a peculiar feature of the Ising model, let us consider 
the other prototypical example for a quantum phase 
transition~\cite{Sachdev99,LSA12}, the Bose-Hubbard model
\begin{equation}
\label{BHH}
\hat H
=
-J
\sum_{i=1}^\ell
\left(
    \hat b_i^\dagger
    \hat b_{i+1}^{\phantom\dagger}
    +
    \hat b_{i+1}^\dagger
    \hat b_i^{\phantom\dagger}
\right)
+
\frac{1}{2}
\sum_{i=1}^\ell
\hat b_i^\dagger
\hat b_i^\dagger
\hat b_{i}^{\phantom\dagger}
\hat b_{i}^{\phantom\dagger}
\,
\end{equation}
that is believed to be non-integrable~\cite{BK03,KB04,HKG09,KRBL10}.
Here $\hat{b}_{i}^\dagger$ and $\hat{b}_{i}$ are the bosonic creation and
annihilation operators at the lattice site $i$. 
As before, we impose periodic boundary conditions. 
Note that the hopping rate $J$ is dimensionless because we measure it in 
units of the on-site interaction energy (usually denoted by $U$).

At unit filling $\langle\hat{n}_i\rangle=1$, there is a quantum phase 
transition (in the thermodynamic limit $\ell\to\infty$)
between the Mott insulator regime where the on-site repulsion 
dominates (in analogy to the paramagnetic state for the Ising model) and 
the superfluid phase where the hopping rate $J$ dominates 
(analogously to the ferromagnetic state). 
Deep in the Mott phase at $J=0$, the ground state factorizes 
$\ket{\Psi}=\bigotimes_i\ket{1}_i$, i.e., it is not entangled.
For increasing $J$, on the other hand, we get correlations such as 
$\langle\hat{b}_{i}^\dagger\hat{b}_{j}\rangle$ which are somewhat analogous to 
the ferromagnetic correlations 
$\langle\hat\sigma_i^z\hat\sigma_j^z\rangle$.

Unfortunately, for the Bose-Hubbard model, entanglement measures in analogy
to the concurrence are not yet available.
There exist genuine bi-partite and multi-partite entanglement measures for 
bosons, but they are known only for special cases such as Gaussian states or 
suitable pure states 
(see~\cite{AFO07,LevayHolweck16} and references therein).
Hence, we focus on the reduced density matrices and their correlated parts. 
We consider a system of finite size (12 bosons on 12 lattice sites) 
and obtain the ground state numerically for arbitrary $J$ by exact 
diagonalization in the subspace of the Hilbert space where the total 
momentum is zero.
This allows to calculate exactly the reduced density matrices.
We find that they contain, in contrast to the Ising model,
in general more than two non-negligible 
eigenvalues, i.e., the approximation~\eqref{approximation} would not 
apply here. 

\begin{figure}[t]

\centering

\includegraphics*[width=.9\linewidth]{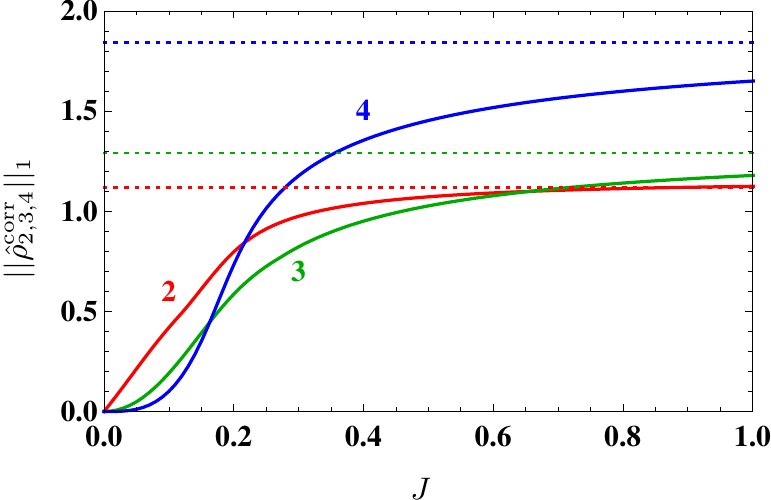}

\caption{(color online) Norms of correlated reduced density operators for 
two $\hat\rho_{2}^{\rm corr}$ (red), 
three $\hat\rho_{3}^{\rm corr}$ (green), and 
four $\hat\rho_{4}^{\rm corr}$ (blue) neighboring sites in the Bose-Hubbard 
model with 12 particles in 12 lattices sites. 
The horizontal dotted lines represent the limit of the ideal Bose 
gas (superfluid phase $J\to\infty$).
We see qualitatively similar features as for the Ising model, except that 
the three-point correlation is monotonically increasing. 
The initial sequence (first two-point, later three-point and even later 
four-point correlations) is also present in this case. 
Again, the four-point correlations overtake the three-point correlations
well before the critical point (here around $J_{\rm crit}\approx0.3$). 
}
\label{bhm_1norms}
\end{figure}

In analogy to Fig.~\ref{norm1.2to4}, we plot the Schatten one-norms of the 
correlated parts of the reduced density matrices in Fig.~\ref{bhm_1norms}.
We find that -- again in contrast to the Ising model -- all three curves are 
monotonically growing and approach finite asymptotic values for $J\to\infty$
which correspond to the limit of a free (ideal) Bose gas and can be 
calculated analytically. 
Similarly to the Ising model, we find 
$||\hat\rho_{q}^{\rm corr}||_1\sim J^{q-1}$ with $q=2,3,4$ for small $J$,
as expected from strong-coupling perturbation theory 
(see Supplement~\cite{Supplement}). 
This scaling imposes the hierarchy 
$||\hat\rho_{2}^{\rm corr}||_1\gg||\hat\rho_{3}^{\rm corr}||_1
\gg||\hat\rho_{4}^{\rm corr}||_1$ at small values of $J$.
However, in analogy to the Ising model, this hierarchy is partially 
inverted at $J\approx 0.16$ and $J\approx 0.21$, i.e. both well before the 
critical point is reached 
(here around  $J_{\rm crit}\approx0.3$, see Ref.~\cite{K16} for a recent review).

\paragraph{Conclusions}
%
For the Ising model~\eqref{Ising}, we study the 
entanglement of two, three, and four neighbouring sites
in the ground state by means of the
approximation~\eqref{approximation} based on the dominance of two eigenvalues. 
We find a sequential increase of entanglement depth with growing $J$
which we call avalanche of entanglement (see Fig.~\ref{tangles}). 
We conjecture that this avalanche continues until pure $\ell$-partite 
(GHZ type) entanglement emerges for $J=\infty$.
This avalanche might also explain the $\ln\ell$ divergence of the entanglement 
entropy at the critical point, which will be subject of future work. 

Using the Schatten one-norms of the correlated reduced density 
matrices as rigorous upper bounds for the correlations~\eqref{Schatten}, 
we find that they also yield approximate upper bounds for the 
corresponding entanglement measures (see Fig.~\ref{tau3-of-rho3}). 
As expected from these observed strong ties between entanglement and 
correlations, we find that the latter display a similar sequence 
(first two-point, later three-point and even later four-point correlations) 
when increasing $J$ (see Fig.~\ref{norm1.2to4}).
However, we also find a partial inversion of the hierarchy of correlations:  
at $J\approx0.8$, i.e., well before the critical point $J_{\rm crit}=1$ is reached, 
the four-point correlations exceed the three-point correlations and eventually also
the two-point correlations. 
Comparison with the Bose-Hubbard model as another prototypical example
reveals a qualitatively similar behavior, including the partial inversion 
of the hierarchy of correlations at $J\approx 0.16$, i.e., well before 
the critical point at $J_{\rm crit}\approx0.3$. 

This inversion of the hierarchy is relevant for approximation 
schemes based on truncation~\cite{Moter92,Fishman92,Jensen94,QNS12,NQS14,QKNS14,KNQS14}.
Let us consider a quantity such as 
$\langle\sigma^x_i \sigma^x_j \sigma^x_k \sigma^x_l \rangle$.
To lowest order (mean-field limit), one could approximate it via 
$\langle\sigma^x_i \sigma^x_j \sigma^x_k \sigma^x_l \rangle
\approx
\langle\sigma^x_i\rangle
\langle \sigma^x_j\rangle
\langle \sigma^x_k\rangle
\langle \sigma^x_l \rangle$,
i.e., by neglecting all correlations.  
As a possible first-order correction, one could include two-point correlations 
such as 
$\langle\sigma^x_i \sigma^x_j\rangle^{\rm corr}
\langle \sigma^x_k\rangle\langle \sigma^x_l \rangle$.
This first-order approximation allows us to derive e.g. the magnon dispersion 
relations. 
One can try to successively improve the accuracy of this approximation by 
shifting the truncation, i.e., by including more and more higher-order 
correlations. 
While this successive approximation procedure works well for small $J$,
we found here that it fails for larger $J$, even well before reaching 
the critical point.
%

\paragraph{Outlook}
%
It might be interesting to study the possibility of more general approximation 
schemes such as~\eqref{approximation} based on the dominance of two or
more eigenvalues of the reduced density operator.
In a time-dependent setting one could analyze how this entanglement avalanche 
is affected by non-adiabatic dynamics during a sweep through the 
critical point.  

\begin{acknowledgments}
This work was supported by the SFB 1242 of the German Research Foundation (DFG).
\end{acknowledgments}



\clearpage


\pagebreak
\begin{widetext}\begin{center}\bf\large%
Supplemental material to\\
Avalanche of entanglement and correlations at quantum phase transitions
\end{center}\end{widetext}
\setcounter{equation}{0}
\setcounter{figure}{0}
\setcounter{table}{0}
\setcounter{page}{1}
\setcounter{section}{1}
\setcounter{enumiv}{1}
\makeatletter
\renewcommand{\theequation}{S\arabic{equation}}
\renewcommand{\thefigure}{S\arabic{figure}}
\renewcommand{\thesection}{S\arabic{section}}
\renewcommand{\theenumiv}{S\arabic{enumiv}}

Here we report on additional material that supports findings of the main paper.

\section{Entanglement measures and correlation functions}
\subsection{Correlated reduced density operators}

We consider the Hamiltonian for a system of $L$ lattice sites of the form
\begin{equation}
\label{Hlat}
\hat H
=
\sum_{\ell_1\ne\ell_2}
\hat H_{\ell_1\ell_2}
+
\sum_{\ell}
\hat H_\ell
\;,
\end{equation}
where $\hat H_\ell$ and $\hat H_{\ell_1\ell_2}$ are local and two-site operators, respectively;
the indices label the lattice sites.
The state of the whole system can be described by the density operator
$\hat\rho=|\psi\rangle\langle\psi|$.
In order to study parts of the system, we introduce reduced density operators 
for $q$ lattice sites via averaging
(partially tracing) over all other sites:
\begin{equation}
\hat\rho_{\ell_1\dots \ell_q}
=
{\rm Tr}_{\ell_{q+1}\dots \ell_L}
\hat\rho
\;,
\end{equation}
where all $\ell_1,\dots ,\ell_L: \{1,\dots,L\}$ are distinct. 
Information about all possible spatial correlations of the lattice sites $\ell_1\dots \ell_q$
is directly contained in the correlated parts of 
the reduced density operator $\hat\rho_{\ell_1\dots \ell_q}^{\rm corr}$.
They are constructed in the same manner as cumulants.
For $q=2,3$ they are explicitly given by
\begin{eqnarray}
\hat\rho_{\ell_1\ell_2}^{\rm corr}
&=&
\hat\rho_{\ell_1\ell_2}-\hat\rho_{\ell_1}\hat\rho_{\ell_2}
\\
\hat\rho_{\ell_1\ell_2\ell_3}^{\rm corr}
&=&
\hat\rho_{\ell_1\ell_2\ell_3}
-
\hat\rho_{\ell_1\ell_2}^{\rm corr}\hat\rho_{\ell_3}
-
\hat\rho_{\ell_1\ell_3}^{\rm corr}\hat\rho_{\ell_2}
-
\hat\rho_{\ell_2\ell_3}^{\rm corr}\hat\rho_{\ell_1}
\nonumber\\
&&-
\hat\rho_{\ell_1}\hat\rho_{\ell_2}\hat\rho_{\ell_3}
\;.
\nonumber
\end{eqnarray}
The operators
$\hat\rho_{\ell_1\dots \ell_q}^{\rm corr}$ are hermitean and their traces vanish:
${\rm Tr}\hat\rho_{\ell_1\dots \ell_q}^{\rm corr}=0$.
They allow to calculate (connected) correlation functions of local operators $\hat O_\ell$ as
\begin{eqnarray}
\langle
   \hat O_{\ell_1} \dots \hat O_{\ell_q}
\rangle^{\rm corr}
=
{\rm Tr}
\left(
    \hat\rho^{\rm corr}_{{\ell_1} \dots {\ell_q}}
    \hat O_{\ell_1} \dots \hat O_{\ell_q}
\right)
\;.
\end{eqnarray}

In order to obtain quantitative estimates of the $q$-point correlations,
it is convenient to consider the Schatten $p$-norms
\begin{equation}
||\hat\rho^{\rm corr}_{{\ell_1} \dots {\ell_q}}||_p
:=
\sqrt[p]{{\rm Tr} |\hat\rho^{\rm corr}_{{\ell_1} \dots {\ell_q}}|^p}
\equiv
\left(
\sum_i |\lambda_{{\ell_1} \dots {\ell_q}}^{(i)}|^p
\right)^{1/p}
\;,
\end{equation}
where $\lambda_{{\ell_1} \dots {\ell_q}}^{(i)}$ are the eigenvalues of the correlated density operators
$\hat\rho^{\rm corr}_{{\ell_1} \dots {\ell_q}}$.
The Schatten one-norm is also known as the trace norm
and the two-norm is often called the Frobenius norm or the Hilbert-Schmidt norm.
Assuming that the aboslute value of the matrix elements
$\langle i|\hat O_{\ell_1}\dots\hat O_{\ell_q}|i\rangle$
in the eigenstates $|i\rangle$ of the operator
$\hat\rho^{\rm corr}_{{\ell_1} \dots {\ell_q}}$ are not larger than one,
it is easy to see that
$
\left|\langle\hat O_{\ell_1} \dots \hat O_{\ell_q}\rangle^{\rm corr}\right|
\le
||\hat\rho^{\rm corr}_{{\ell_1} \dots {\ell_q}}||_1
$.

In the present work, we deal with the ground states of one-dimensional 
translationally invariant systems.
In this case, the density matrices depend only on the distances between the lattices sites.
One can always order the site indices such that $\ell_1<\ell_2<\dots<\ell_q$
and we can write
$\hat\rho^{\rm corr}_{{\ell_1} \dots {\ell_q}} \equiv \hat\rho_q^{\rm corr}(d_1,\dots,d_{q-1})$,
where $d_i=\ell_{i+1}-\ell_i>0$ are the corresponding distances.
Intuitively, one would expect that
(i) the correlations of a fixed number of sites decrease with the distances between the sites and
(ii) the correlations for fixed distances decrease with the number of sites.
For nearest neighbors, the latter would lead to inequalities
\begin{equation}
\label{ineq}
\!
||\hat\rho^{\rm corr}_2(1)||_p
\gg
||\hat\rho^{\rm corr}_3(1,1)||_p
\gg
\dots
\end{equation}
Our analysis of two completely different models presented below shows that
the expectation (i) is always satisfied whereas (ii) does not necessarily hold.

\subsection{Correlations as upper bounds to entanglement}

\begin{figure}[t]
\centering

\includegraphics*[width=.9\linewidth]{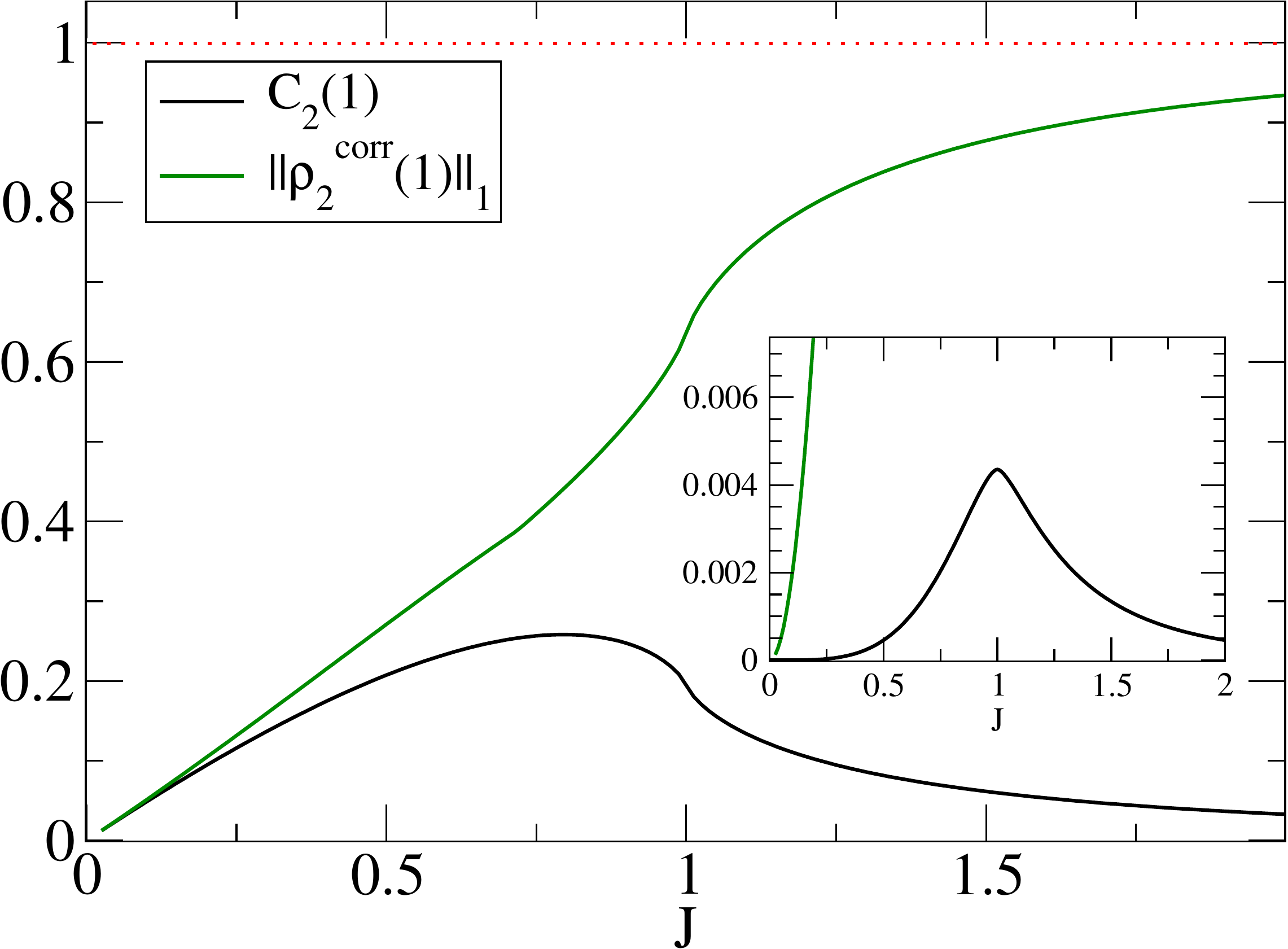}

\caption{The concurrence $C_2(d)$ is plotted together with 
$|\!|\rho_2^{\rm corr}(d)|\!|_1$ for distances $d=1$ (left figure) 
and $d=2$ (right figure). For $d=1$, it is clearly seen that indeed 
$|\!|\rho_2^{\rm corr}(1)|\!|_1$ as largest correlation function
is an upper bound to the corresponding concurrence. For $d=2$,
the concurrence drops to $0.04$ at its maximum at the critical point, wheras 
$|\!|\rho_2^{\rm corr}(2)|\!|_1$ is about roughly the same as
$|\!|\rho_2^{\rm corr}(1)|\!|_1$. }

\label{Corr.1u2}
\end{figure}

For pure states, the largest correlation function
$||\hat\rho_{\ell_1\ell_2}^{\rm corr}||_1$
coincides with the concurrence $C_2(\ell_1,\ell_2)$~\cite{sVerstraetePC04}. 
This strict equality for pure states
turns into an upper bound for the concurrence of mixed states 
(see Fig.~\ref{Corr.1u2}).
It is interesting to see whether the respective correlation functions 
are an upper bound to the corresponding entanglement measure.

\begin{figure}[t]
\centering

\includegraphics*[width=.9\linewidth]{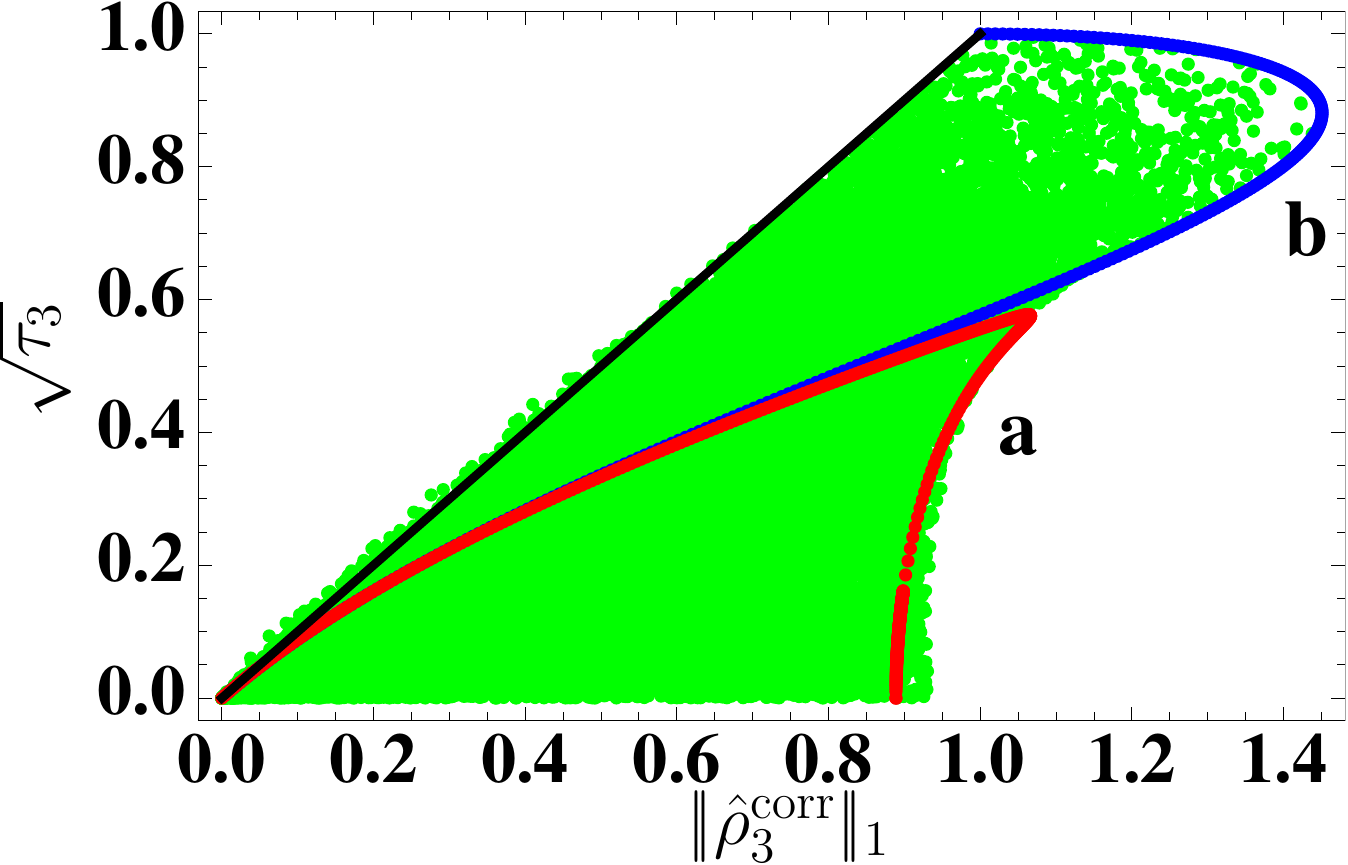}

\caption{The three-tangle is plotted against the largest  correlation
for pure states. We have done a plot for each state as labled by an Acin state~\cite{sAcin01}
statistically. Each dot in the figure corresponds 
to a single state. Although the largest  correlation function 
is not always larger that the three-tangle (as for two qubits) it is 
however satisfied approximately (see the thin black line, signalling 
equality of the both). The green and red curve are GHZ states which 
are made of two and four components, respectively. The four component GHZ state
connects the W states with the product states. It is seen that 
some states do exist with an even larger $|\!|\rho_3^{\rm corr}|\!|_1$  
as the W state. }

\label{tau3-of-rho3_s}
\end{figure}

For the three-tangle $\tau_3$ we find that it is almost upper bounded by
$|\!|\hat\rho_3^{\rm corr}|\!|_1$ for pure states of three qubits
(see Fig.~\ref{tau3-of-rho3_s} or Fig.~\ref{tau3-of-rho3} in the main paper). 
There are however some pure states
for which $\tau_3$ is slighly above  $|\!|\hat\rho_3^{\rm corr}|\!|_1$.

\begin{figure}
\centering

\includegraphics*[width=.9\linewidth]{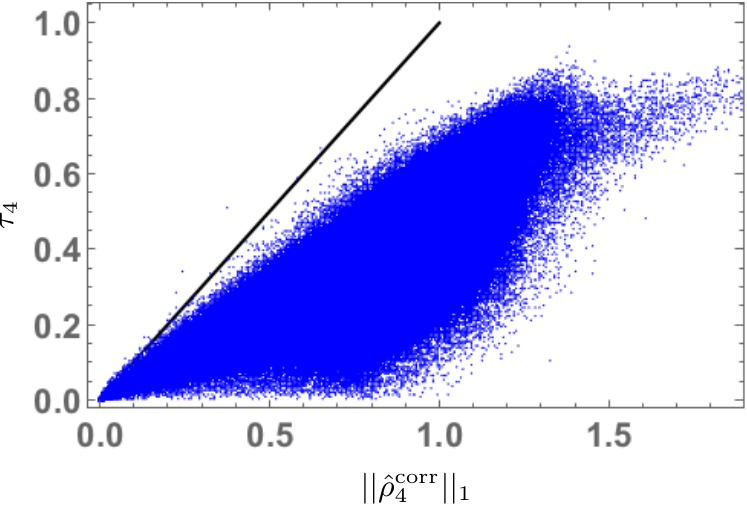}

\caption{The four-tangle $\tau_4$ is shown here against $|\!|\hat\rho_4^{\rm corr}|\!|_1$ for pure states.
We have chosen the pure states statistically from the extended Schmidt form~\cite{sCarteret00}.
Each dot in the figure corresponds to a single state. We do not have enough statistics, as can be seen from 
our viewgraph. Although states for arbitrary value of $\tau_4$ should be there whose value 
of $|\!|\hat\rho_4^{\rm corr}|\!|_1$ comes arbitrarily close to the line $\tau_4=|\!|\hat\rho_4^{\rm corr}|\!|_1$,
we don't see any occurrences at larger values of $\tau_4$.}

\label{tau4-of-rho4}
\end{figure}

We do not reach a satisfactory statistics to make a similar statement also 
in the situation of pure states for four qubits. The result is shown for
900.000 random choices of pure states for $\tau_4^{a}$ 
in Fig.~\ref{tau4-of-rho4}. It can be seen, however, that for almost all states
out of this sample the inequality holds; there are examples shown for which 
$|\!|\hat\rho_4^{\rm corr}|\!|_1$ is smaller than $\tau_4^{a}$.

The genuine multipartite entanglement content, i.e. the three-tangle
$\sqrt{\tau_3}$ and the three four-tangles $\tau_4^a=\sqrt[3]{{\cal F}_1^{(4)}}$, 
$\tau_4^b=\sqrt[4]{\left\langle{\cal F}_2^{(4)}\right\rangle_s}$, and 
$\tau_4^c=\sqrt[6]{{\cal F}_3^{(4)}}$ from Ref.~\cite{sDoOs08}
for the nearest-neighboring sites are shown in Fig.~\ref{tangles}
of the main article.
Numerical calculations show that $\tau_4^a$, $\tau_4^b$, $\tau_4^c$ 
for the nearest neighbors are the same,
although we do not have a rigorous analytical proof of that.
Therefore, we do not need to distinguish between the 
three four-tangles and drop the upper index.
This also means that the entanglement would only be due to the 
three- or four-particle GHZ-states, respectively.
Such a behaviour goes conform with the expectations 
for that particular model~\cite{OstUpperBound16}.

\section{Results for the Ising model}

\subsection{Rank-two approximation}

\begin{figure}
\centering

\includegraphics*[width=.9\linewidth]{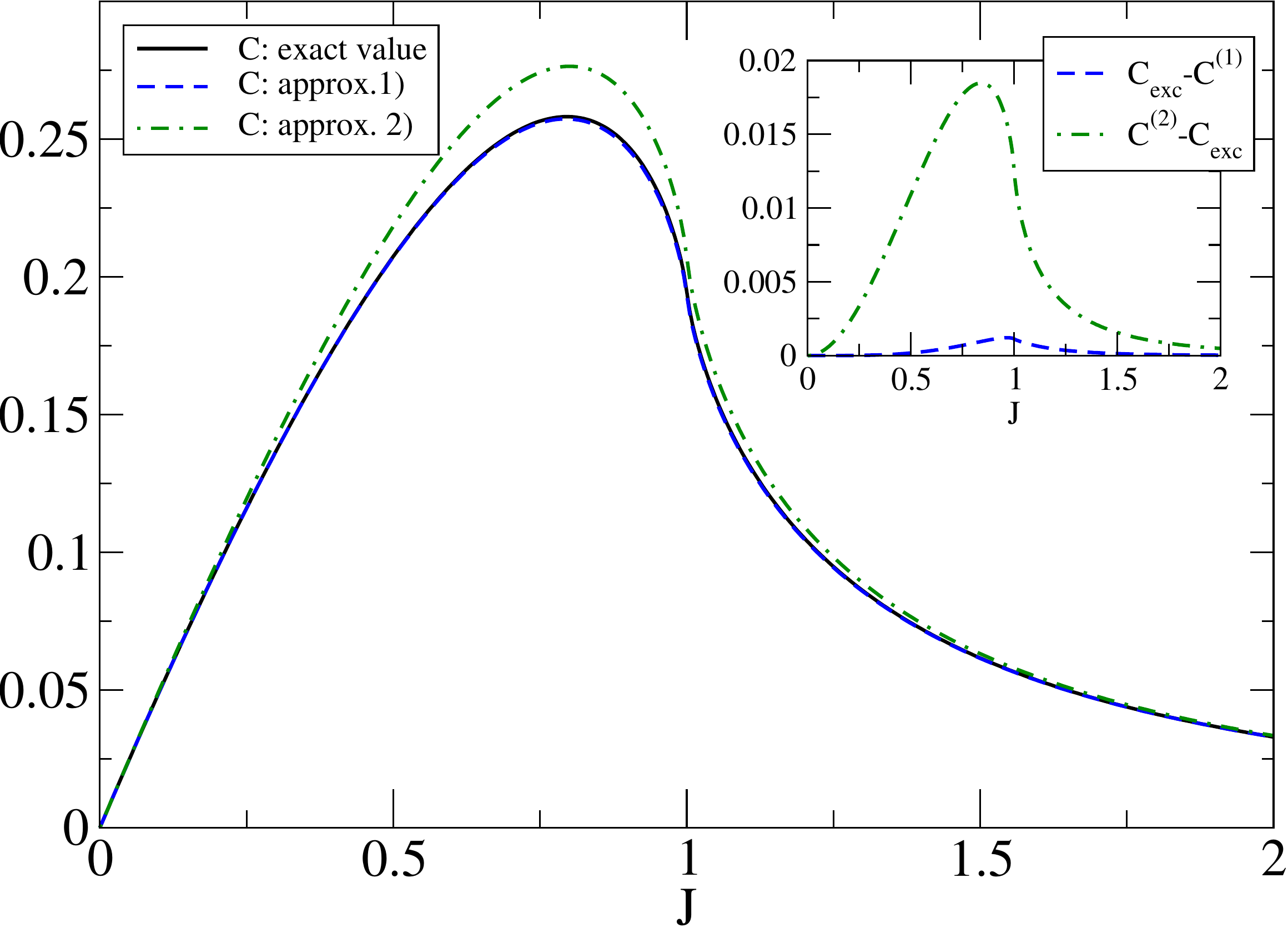}

\includegraphics*[width=.9\linewidth]{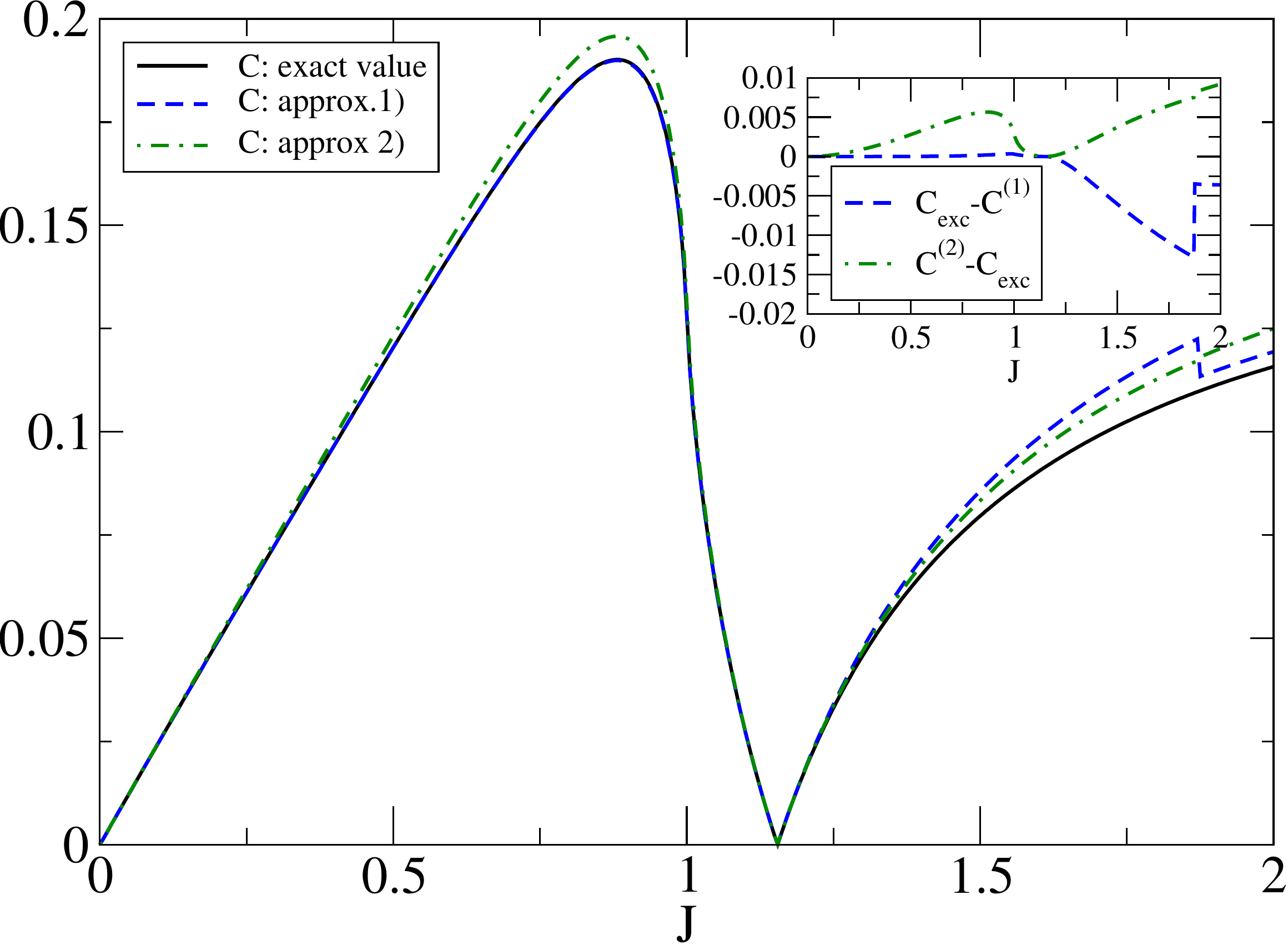}

\caption{
Top:
The concurence $C_2(1)$ is plotted together with the two approximations 1) and 2) (see text)
for the transverse Ising model. 
It can be seen that approximation 1) (blue dashed curve; almost invisible here) 
basically coincides with the exact concurrence (black curve). The approximation following scheme 2)
(green dash-dotted curve) slightly lies above the exact curve. This is demonstrated in the inset, where the 
differences $C_{{\rm Diff.}}^{(1)}:=C-C^{(1)}$ (blue dashed curve) and $C_{{\rm Diff.}}^{(1)}:=C^{(2)}-C$ (red curve) are shown.
\\
Bottom:
The same plots as for the transverse Ising chain are shown here for the transverse XY model
with anisotropy parameter $\gamma=0.5$. Here the concurrence $C_2(1)$ 
is well described by approximation scheme 1) up to the factorising field. Beyond this point, 
the approximation is still reasonable, but lies above the exact curve.}

\label{Conc}
\end{figure}

\begin{figure}
\centering

\includegraphics*[width=.9\linewidth]{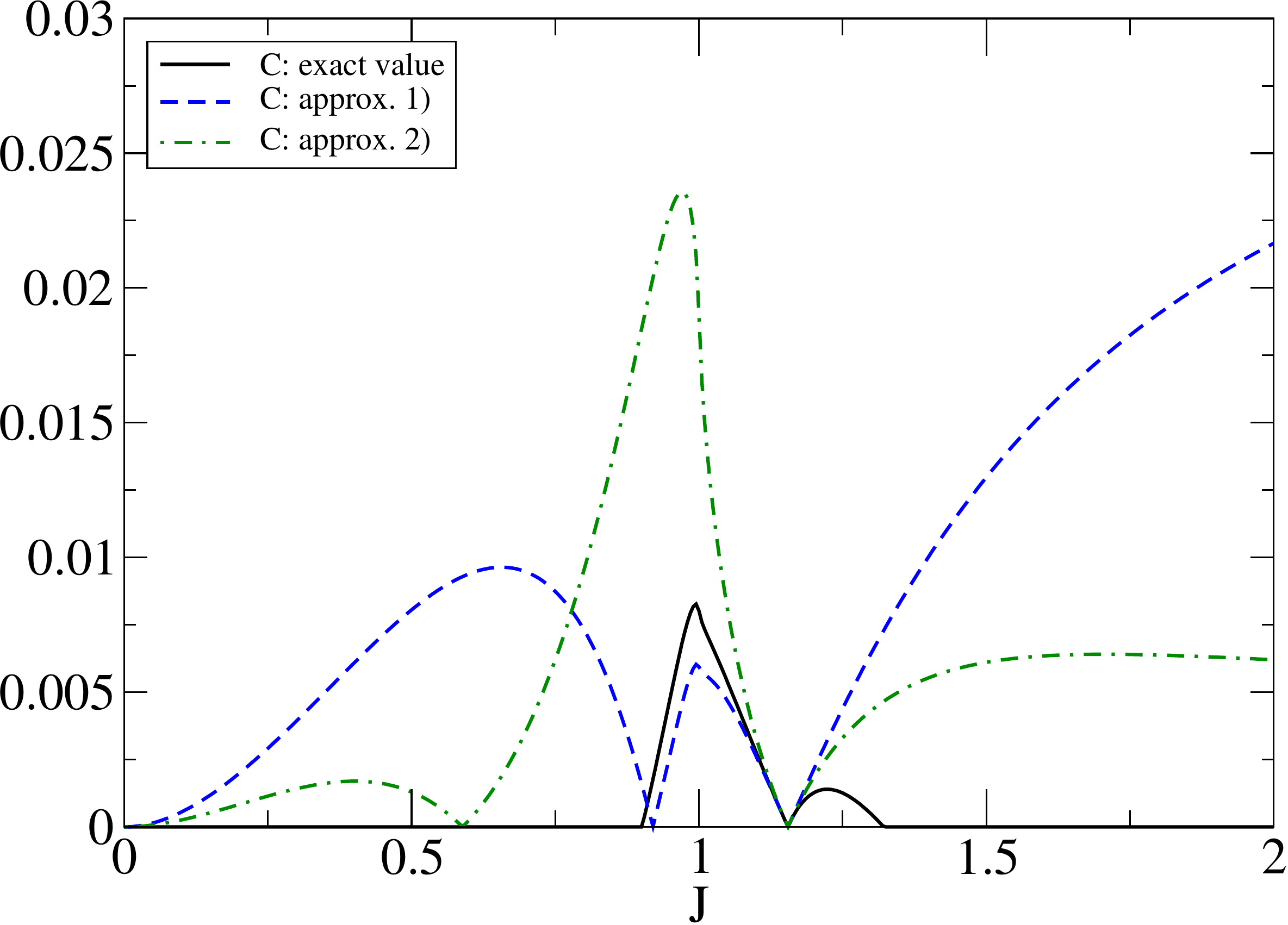}

\caption{The concurrence $C_2(3)$ is shown for the transverse XY-model and anisotropy parameter $\gamma=0.5$.
The approximation following scheme 1) is reasonable around the critical point in between the two zeros.
Beyond these points it deviates considerably from zero. 
It however gives a close prediction of the non-trivial zero of $C_2(3)$.}

\label{Concd3g0-5}
\end{figure}

The quantum transverse Ising model is a special case of the transverse XY-model
\begin{equation}
\label{XY}
\hat H
=
-J\sum_{i=1}^\ell\left(\frac{1+\gamma}{2}\hat\sigma_i^x\hat\sigma_{i+1}^x 
+\frac{1-\gamma}{2}\hat\sigma_i^y\hat\sigma_{i+1}^y\right)
-\sum_{i=1}^\ell\hat\sigma_i^z
\,.
\end{equation}
For $\gamma\neq 0$ it has a quantum phase transition of the Ising type.
The reduced density matrices of two, 
$\hat\rho_2=\hat\rho_{ij}$, three, $\hat\rho_3=\hat\rho_{ijk}$, and four, 
$\hat\rho_4=\hat\rho_{ijkl}$, neighboring spins~\cite{sPFEUTY,sHofmann14}
essentially possess two
dominant eigenvalues $p_1$ and $p_2$ while the sum of the remaining 
sub-dominant eigenvalues stays below $2.5\%$.
The second eigenvector interferes strongly with the entanglement of the first, 
whereas we checked that this is not the case for the third; 
the remaining error is maximally $p_{I>3}\approx 0.5\%$. 
Therefore we only consider the case of rank two density matrices
and neglect the rest of few percents of weight. 
What one is left with are the two 
highest weights $p_1$ and $p_2$ of the density matrix. 
We briefly discuss two ways of taking care of them: 
1) take $p_1$ 
or 
2) take $p_1/(p_1+p_2)$
as the highest weight of the new rank-two density matrix. 
Whereas in 1) one assumes that the neglected part 
is as destructive to the entanglement as the second state is, in 2) the remaining states do not enter
the calculation at all.
Both are neither an upper bound nor a lower bound to the entanglement.
Details on how the approximations works for the concurrence can be seen in Figs.~\ref{Conc} and \ref{Concd3g0-5}.

The procedure 1) works perfectly for nearest neighbors, where the plots can hardly be distinguished
for the transverse Ising model and also for the transverse XY-model for $J$ 
up to the factorising field, where they start to deviate considerably.
This is different when considering larger distances, where both curves have similar shapes 
only around the critical point (see Fig.~\ref{Concd3g0-5}). 
It gives good estimates even for the zeros of the exact concurrence
and avoids over-estimating the entanglement in the state.

\subsection{Correlation functions}

The concurrences $C_2(\ell_1,\ell_2)\equiv C_2(d)$, where $d=|\ell_2-\ell_1|$ 
are shown together with the 1-norm of $\rho_2^{\rm corr}$ in Fig.~\ref{Corr.1u2}. 
Wheras the 1-norm of the correlations has no substantial
changes, the concurrence at distance $d=2$ modifies to about $2\%$ of the maximal value 
for nearest neighbors (see inset).

\begin{figure}[t]
\centering


\includegraphics*[width=.9\linewidth]{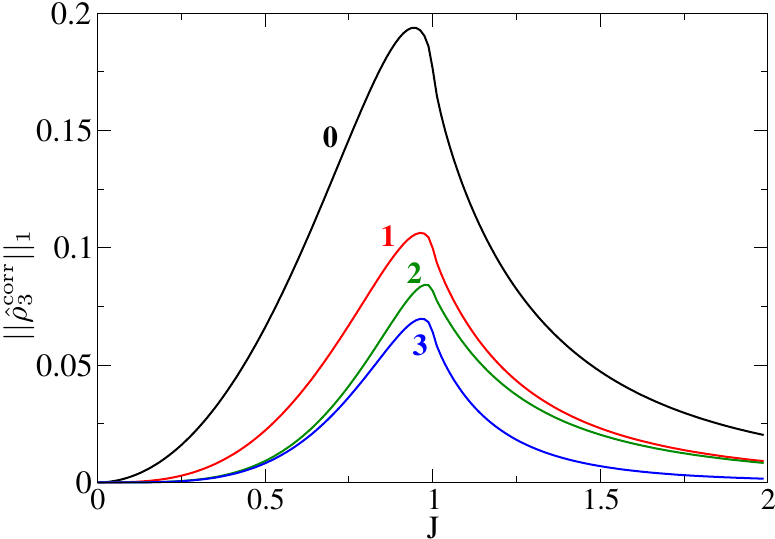}

\caption
{
$|\!|\hat\rho_3^{\rm corr}(d_1,d_2)|\!|_1$ is shown for $(d_1,d_2)$ from nearest neighbors $(1,1)$~(black, 0)
to $(1,3)$~(green, 2) together with $(2,2)$~(blue, 3). It is not much affected like for the two-site case,
in that it only goes down to about one third of $|\!|\hat\rho_3^{\rm corr}(1,1)|\!|_1$. The maximum is
assumed at roughly $J\gtrsim 0.945$ and is slighly moving towards the critical point $J_c=1$
when the sites are moving away from each other.
}

\label{Corr3}
\end{figure}

\begin{figure}[t]
\centering


\includegraphics*[width=.9\linewidth]{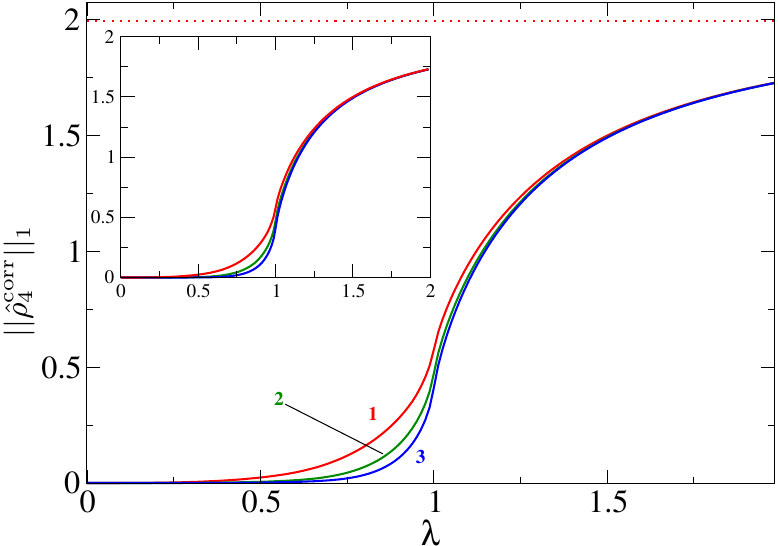}

\caption{$|\!|\hat\rho_4^{\rm corr}|\!|_1$ is shown for distances corresponding to $(1,d_2,1)$ for 
$d=1$ (red, 1) to $3$ (blue, 3). The effect of growing $d$ seems to be that the curve for $d$ be an upper limit to $m$ with $m<n$.
The major change is done around the critical value of $J_c=1$.
The inset shows the distances $(2,1,2)$ and $(2,2,2)$ as compared to $(1,1,1)$.
There is not much difference noted.}

\label{rho4.1-n-1.gamma1}
\end{figure}

This qualitatively doesn't change much for $3$ sites at distances $(d_1,d_2)$ which means that if
the first particle is at site $\ell_1$ the next site is at $\ell_1+d_1$ and the third one at $\ell_1+d_1+d_2$
(hence, the next neighbor reduced density matrix would be $\hat\rho_3(1,1)$). 
This is shown in Fig.~\ref{Corr3},
where different distances have been considered for $|\!|\hat\rho^{\rm corr}(d_1,d_2)|\!|_1$:
$(d_1,d_2)=(1,1)$ to $(1,3)$ and $(2,2)$. $|\!|\hat\rho_3^{\rm corr}(d_1,d_2)|\!|_1$ decays to about one third 
of the nearest neighbor situation $|\!|\hat\rho_3^{\rm corr}(1,1)|\!|_1$ with a maximum at $J\gtrsim 0.945$
which is moving tinily up to $0.98$. One could extract the tendency that the maximum moves 
for $(1,1)$ to $(1,d_2)$ and from $(1,1)$ to $(d_1,d_1)$ closer to the critical point (where in the ulimate example we have only 
considered the additional case $(2,2)$).
Observe the astonishingly parallel situation to the two-site case,
namely that $|\!|\hat\rho_3^{\rm corr}(d_1,d_2)|\!|_1$ doesn't change so drastically with growing distance.
This continues to hold for $|\!|\hat\rho_4^{\rm corr}(d_1,d_2,d_3)|\!|_1$
(see Fig.~\ref{rho4.1-n-1.gamma1}).

\begin{figure}[t]

\centering

\includegraphics*[width=.9\linewidth]{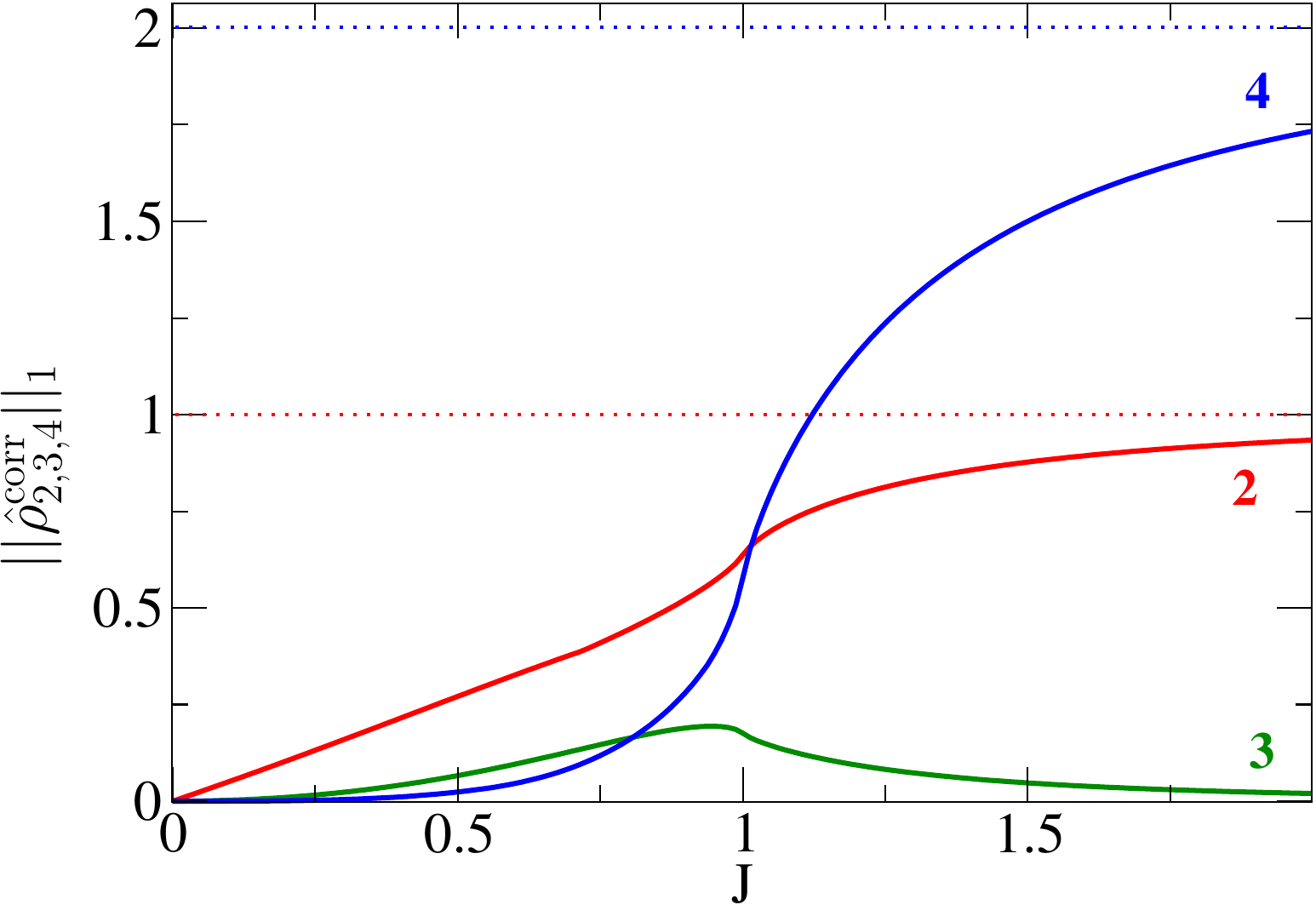}

\includegraphics*[width=.9\linewidth]{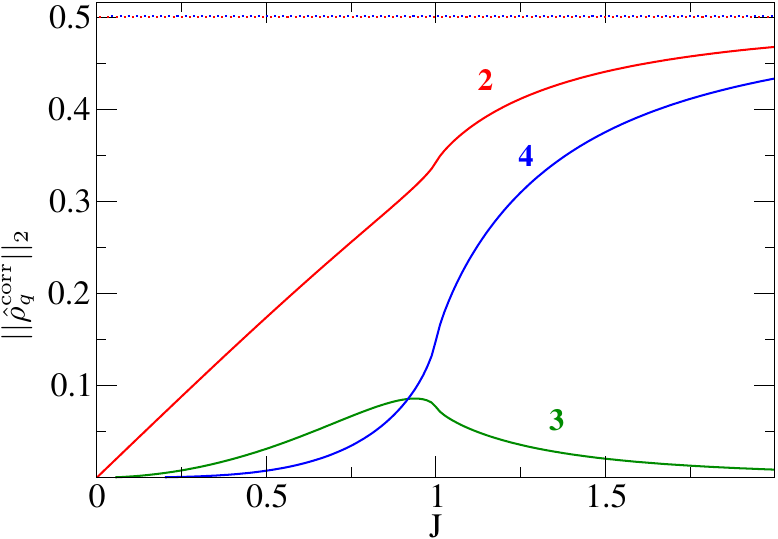}

\caption{
         Norms of correlated reduced density operators
         $\hat\rho_{2}^{\rm corr}(1)$~(red,2),
         $\hat\rho_{3}^{\rm corr}(1,1)$~(green,3),
         $\hat\rho_{4}^{\rm corr}(1,1,1)$~(blue,4)
         for the ground state of the transverse Ising model.
         (a)~1-norms.
         For $J\lesssim0.8$,
         $||\hat\rho^{\rm corr}_2(1)||_1>||\hat\rho^{\rm corr}_3(1,1)||_1>||\hat\rho^{\rm corr}_4(1,1,1)||_1$.
         For $0.8\lesssim J\lesssim1$,
         $||\hat\rho^{\rm corr}_2(1)||_1>||\hat\rho^{\rm corr}_4(1,1,1)||_1>||\hat\rho^{\rm corr}_3(1,1)||_1$.
         For $J\gtrsim1$,
         $||\hat\rho^{\rm corr}_4(1,1,1)||_1>||\hat\rho^{\rm corr}_2(1)||_1>||\hat\rho^{\rm corr}_3(1,1)||_1$.
         (b)~\mbox{2-norms}.
         $||\hat\rho^{\rm corr}_2(1)||_2$ is always larger than $||\hat\rho^{\rm corr}_3(1,1)||_2$ and $||\hat\rho^{\rm corr}_4(1,1,1)||_2$.
         For $J\lesssim0.9$, $||\hat\rho^{\rm corr}_4(1,1,1)||_2 < ||\hat\rho^{\rm corr}_3(1,1)||_2$,
         and for $J\gtrsim0.9$ we have the opposite.
        }

\label{norm1.2to4_s}
\end{figure}

\begin{figure}[t]

\centering

\includegraphics*[width=.9\linewidth]{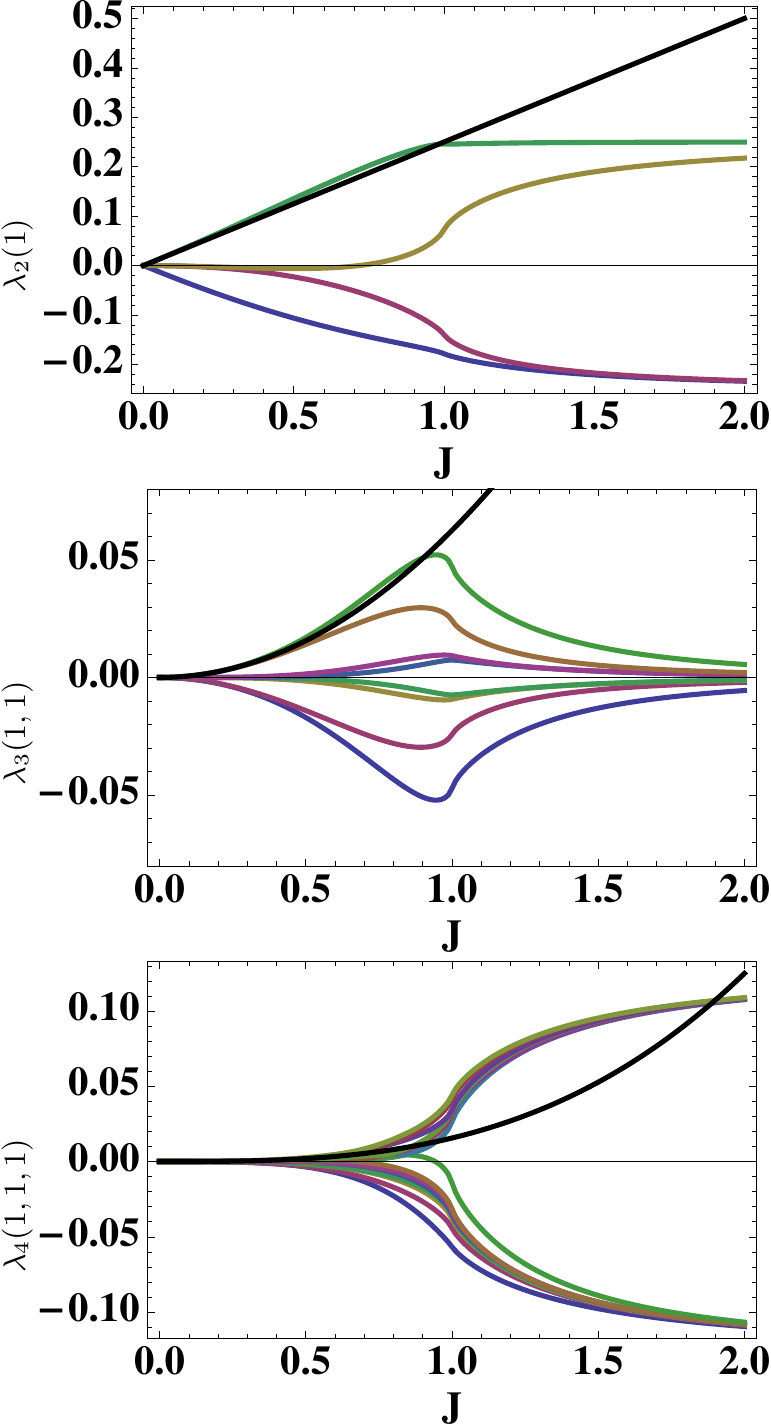}

\caption{
         Eigenvalues of the correlated reduced density operators
         $\hat\rho_{2}^{\rm corr}(1)$~(a),
         $\hat\rho_{3}^{\rm corr}(1,1)$~(b),
         $\hat\rho_{4}^{\rm corr}(1,1,1)$~(c)
         for the ground state of the transverse Ising model.
         }

\label{rhoconnevals.2to4}
\end{figure}

\subsection{Difference in the norm}

We have seen that the 1-norm serves as an upper bound to
the correlation functions in the model. We nevertheless studied also the 
2-norm, known as the Hibert-Schmidt or Frobenius norm.
The result is shown in Fig~\ref{norm1.2to4_s}.
It is seen that the 2-norm close to the crtitical point
$\hat\rho_{4}^{\rm corr}(1,1,1)$ is still 
considerably larger than $\hat\rho_{3}^{\rm corr}(1,1)$; 
it is however still smaller than $\hat\rho_{2}^{\rm corr}(1)$ 
showing only a partial reordering. The corresponding eigenvalues of
the matrices $\hat\rho_{q}^{\rm corr}(1)$, $q=2,3,4$ is shown in
Fig.~\ref{rhoconnevals.2to4}. 

\section{Results for the Bose-Hubbard model}

The ground state of the Hamiltonian~(\ref{BHH}) is obtained numerically for arbitrary $J$
by exact diagonalization in the subspace of the Hilbert space where the total momentum is zero.
This allows to calculate exactly the reduced density matrices.
In the basis of the occupation numbers $n_1\dots n_q$,
the entries $\langle n_1\dots n_q | \hat\rho_q(d_1,\dots,d_{q-1}) | n_1'\dots n_q'\rangle$
do not vanish, provided that
$
\sum_{i=1}^q
n_i
=
\sum_{i=1}^q
n_i'
=
n_{\rm B}
=0,\dots,N
$.
Thus, the reduced density matrices possess a block-diagonal structure and the blocks are labeled by $n_{\rm B}$.
The correlated density matrices have a similar structure.

\begin{figure}[t]

\includegraphics*[width=.9\linewidth]{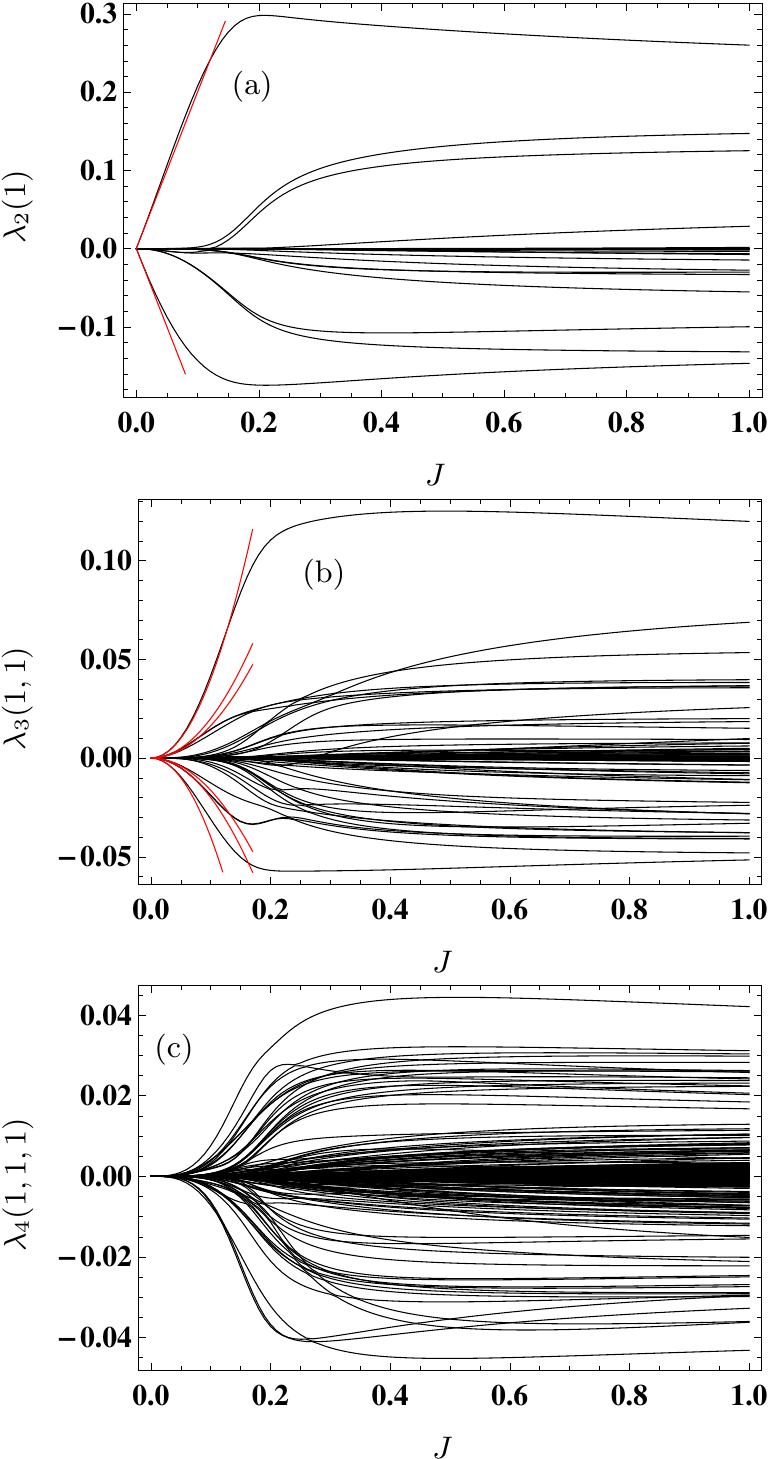}

\caption{
         Eigenvalues of the correlated reduced density operators
         $\hat\rho_{2}^{\rm corr}(1)$~(a),
         $\hat\rho_{3}^{\rm corr}(1,1)$~(b),
         $\hat\rho_{4}^{\rm corr}(1,1,1)$~(c)
         for the ground state of the Bose-Hubbard model.
         Black lines -- exact diagonalization for $N=L=12$.
         Red lines -- strong-coupling expansion, see Eqs.~(\ref{evcorr}).
        }
\label{ev}
\end{figure}

\begin{figure}[t]

\includegraphics[width=.9\linewidth]{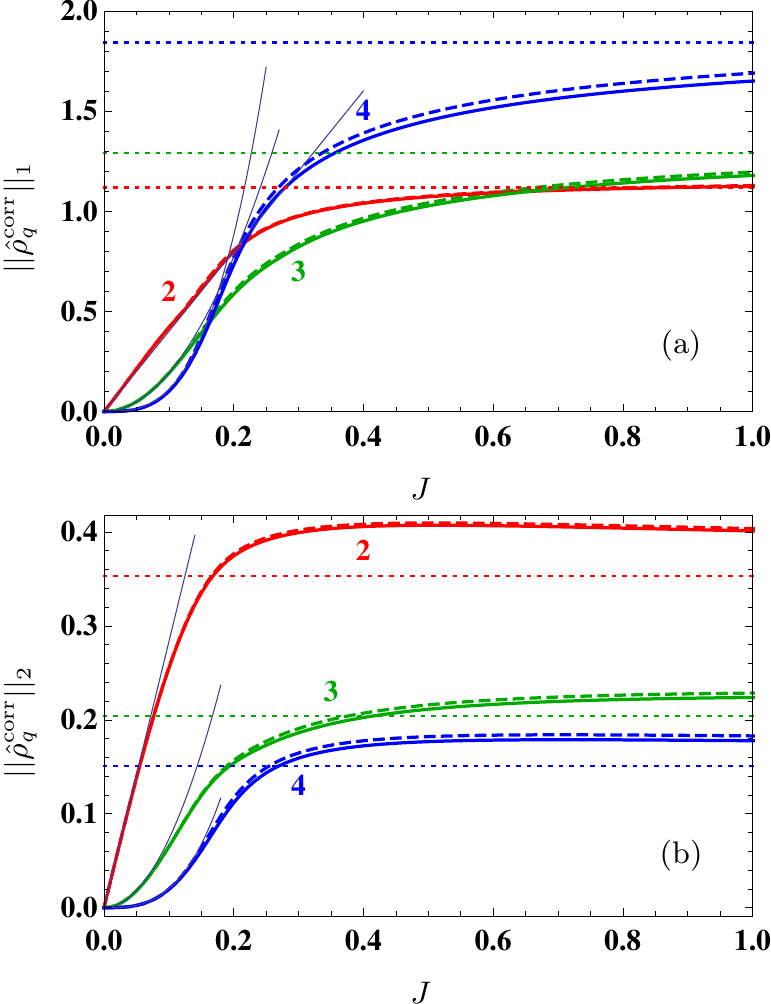}

\caption{
         Norms of correlated reduced density operators
         $\hat\rho_{2}^{\rm corr}(1)$~(red,2),
         $\hat\rho_{3}^{\rm corr}(1,1)$~(green,3),
         $\hat\rho_{4}^{\rm corr}(1,1,1)$~(blue,4)
         for the ground state of the Bose-Hubbard model.
         The results of exact diagonalization are shown by dashed curves for $N=L=9$
         and by solid curves for $N=L=12$.
         Horizontal dotted lines - the limit of the ideal Bose gas ($J\to\infty$, $N=L=12$).
         Thin solid lines -- strong-coupling expansion [Eq.~(\ref{norms_sce1})].
        }
\label{bhm_norms}
\end{figure}

The eigenvalues of the correlated reduced density operators $\hat\rho_q^{\rm corr}(1,\dots,1)$ are shown in Fig.~\ref{ev}.
With the increase of the number of lattice sites $q$, 
the number of nonvanishing eigenvalues grow but their magnitudes decrease.
This leads to a qualitatively different behavior of the one- and two-norms 
that are plotted in Fig.~\ref{bhm_norms}.
The one-norms $||\hat\rho_q^{\rm corr}(1,\dots,1)||_1$ grow monotonically with the increase of $J$ and tend
to finite constant values in the limit $J\to\infty$.
For small values of $J$, we indeed have~(\ref{ineq})
but already at moderate values of $J$ the one-norms for different $q$ become comparable to each other.
It is quite surprising that $||\hat\rho^{\rm corr}_4(1,1,1)||_1$ becomes quickly larger than
$||\hat\rho^{\rm corr}_3(1,1)||_1$ and later also larger than $||\hat\rho^{\rm corr}_2(1)||_1$. 
This happens much before the critical point $J_{\rm c}$ of the superfluid--Mott-insulator transition.
Hence we observe the same behavior as for the integrable Ising model.

The two-norms $||\hat\rho_q^{\rm corr}(1,\dots,1)||_2$ display completely different behavior because
the contribution of small eigenvalues is suppressed. 
The inequalities~(\ref{ineq}) are satisfied for the two-norms at any value of $J$,
although the difference between $q=2,3,4$ is not very large near and above $J_{\rm c}$.
The two-norms possess broad maxima and approach their asymptotic values at $J\to\infty$ from above.

\begin{figure}[t]

\includegraphics*[width=.9\linewidth]{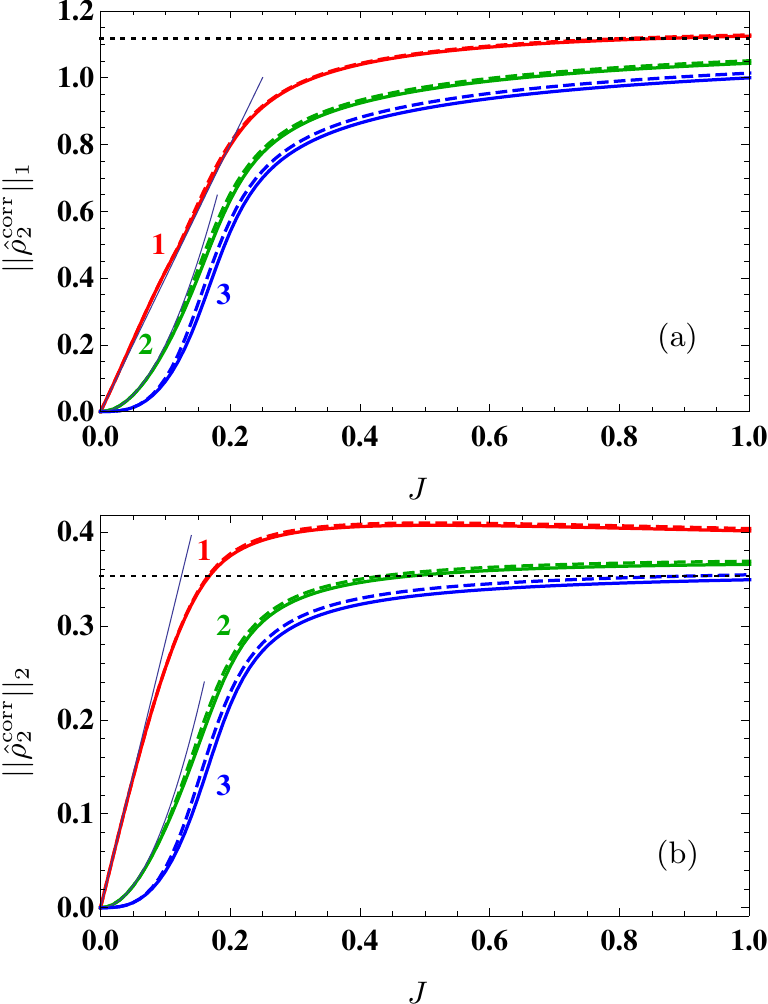}

\caption{
         Norms of correlated reduced density operators
         $\hat\rho_{2}^{\rm corr}(1)$~(red,1),
         $\hat\rho_{2}^{\rm corr}(2)$~(green,2),
         $\hat\rho_{2}^{\rm corr}(3)$~(blue,3)
         for the ground state of the Bose-Hubbard model.
         The results of exact diagonalization are shown by dashed curves for $N=L=9$
         and by solid curves for $N=L=12$.
         Horizontal dotted lines - the limit of the ideal Bose gas ($J\to\infty$, $N=L=12$).
         Thin solid lines -- strong-coupling expansion [Eq.~(\ref{norms_sce_n})].
        }
\label{bhm_norms_2p}
\end{figure}

If we consider one- and two-norms for fixed $q$ but vary the distances between the sites,
we find that both norms decrease with the distance which is demonstrated in Fig.~\ref{bhm_norms_2p} for two sites ($q=2$).
The same was also observed for the transverse Ising model.

The correlated reduced density matrices can be calculated analytically for small values of $J$, 
employing the strong-coupling expansion~\cite{sFM96}.
In the leading order of $J$, this gives the following results for their nonvanishing eigenvalues
\begin{eqnarray}
\label{evcorr}
\lambda_2^{(\pm1)}(1)
&\approx&
\pm
\sqrt{2n(n+1)}
\,J
\;,
\\
\lambda_2^{(\pm1)}(2)
&=&
\lambda_2^{(\pm2)}(2)
\approx
\pm
n(n+1)
\,J^2
\;,
\nonumber\\
\lambda_2^{(\pm2)}(2)
&\approx&
\pm
(2n+1)
\sqrt{2n(n+1)}
\,J^2
\;,
\nonumber\\
\lambda_3^{(\pm1)}(1,1)
&\approx&
\pm
2n(n+1)
\,J^2
\;,
\nonumber\\
\lambda_3^{(\pm2)}(1,1)
&=&
r_3^{(\pm3)}(1,1)
\approx
\pm
n(n+1)
\,J^2
\;,
\nonumber\\
\lambda_3^{(\pm4)}(1,1)
&\approx&
\pm
\frac{2}{3}
\sqrt{n(n+1)(2n^2+2n-1)}
\,J^2
\;,
\nonumber
\end{eqnarray}
where $n=N/L$ is assumed to be an arbitrary integer.
Then for the norms we get
\begin{eqnarray}
\label{norms_sce_n}
||\hat\rho_2^{\rm corr}(1)||_1
&\approx&
2 \sqrt{2n(n+1)}
\;
J
\;,
\\
||\hat\rho_2^{\rm corr}(1)||_2
&\approx&
2 \sqrt{n(n+1)}
\;
J
\;,
\nonumber\\
||\hat\rho_2^{\rm corr}(2)||_1
&\approx&
2
\left[\vrule width0em height2.58ex depth0ex
    2n(n+1)\right.
    \nonumber\\
 &&\left. \ +   (2n+1)\sqrt{2n(n+1)}
\right]
J^2
\;,
\nonumber\\
||\hat\rho_2^{\rm corr}(2)||_2
&\approx&
2
\sqrt{n(n+1)(5n^2+5n+1)}
\,J^2
\;,
\nonumber\\
||\hat\rho_3^{\rm corr}(1,1)||_1
&\approx&
\left[\vrule width0em height3.58ex depth0ex
    8 n(n+1)\right.\nonumber\\
&& \left.\   +
    \frac{4}{3}
    \sqrt{n(n+1)(2n^2+2n-1)}
\right]
J^2
\;,
\nonumber\\
||\hat\rho_3^{\rm corr}(1,1)||_2
&\approx&
\frac{2}{3}
\left\{
    n(n+1)
    \left[
        31n(n+1)-2
    \right]
\right\}^{1/2}
J^2
\;.
\nonumber
\end{eqnarray}
In the special case $n=1$, this gives
\begin{eqnarray}
\label{norms_sce1}
&&
||\hat\rho_2^{\rm corr}(1)||_1 \approx 4 J
\;,\quad
||\hat\rho_2^{\rm corr}(1)||_2 \approx 2.82 J
\;,
\\
&&
||\hat\rho_3^{\rm corr}(1,1)||_1 \approx 19.266 J^2
\;,\quad
||\hat\rho_3^{\rm corr}(1,1)||_2 \approx 7.3 J^2
\;,
\nonumber
\end{eqnarray}
which is in excellent agreement with our numerical calculations (see Fig.~\ref{bhm_norms}).

In the limit of the ideal Bose gas ($J\to\infty$), the entries of the reduced density matrices
depend only on the number of sites $q$ but not on the distances between those:
\begin{eqnarray}
\label{rdm_ibg}
&&
\!\!\!\!\!\langle n_1 \dots n_q |
\hat\rho_q
| n_1' \dots n_q' \rangle
=
\frac{N!}{(N-n_{\rm B})!n_{\rm B}!}\left(
    \frac{n_{\rm B}!}{n_1!\dots n_q!}
\right)^{1/2}
\nonumber\\
&&
\qquad
 \times
\left(
    \frac{n_{\rm B}!}{n_1'!\dots n_q'!}
\right)^{1/2}
\left(
    1 - \frac{q}{L}
\right)^{N-n_{\rm B}}
\left(
    \frac{1}{L}
\right)^{n_{\rm B}}
\;.
\end{eqnarray}
Eq.~(\ref{rdm_ibg}) leads to rather simple expressions for the 2-norms in the thermodynamic limit
\begin{eqnarray}
||\hat\rho_2^{\rm corr}||_2
&=&
\left[
    I_0(4\langle\hat n_\ell\rangle)
    -
    I_0^2(2\langle\hat n_\ell\rangle)
\right]^{1/2}
e^{-2\langle\hat n_\ell\rangle}
\;,
\nonumber\\
||\hat\rho_3^{\rm corr}||_2
&=&
\left[
    I_0(6\langle\hat n_\ell\rangle)
    -
    3 I_0(2\langle\hat n_\ell\rangle) I_0(4\langle\hat n_\ell\rangle)
\right.
\nonumber\\
    &&+
\left.
    2 I_0^3(2\langle\hat n_\ell\rangle)
\right]^{1/2}
e^{-3\langle\hat n_\ell\rangle}
\;,
\nonumber
\end{eqnarray}
where $\langle\hat n_\ell\rangle=N/L$ is not necessarily an integer and
$I_0(x)$ is the modified Bessel function of the first kind.
For $\langle\hat n_\ell\rangle=1$, this yields
$||\hat\rho_2^{\rm corr}||_2\approx 0.334$, $||\hat\rho_3^{\rm corr}||_2\approx 0.184$.
These values are slightly lower than those shown in Fig.~\ref{bhm_norms}(b) indicated by the horizontal dotted lines,
which is a manifestation of the finite-size effects.




\begin{thebibliography}{48}
\bibitem[{foo({\natexlab{a}})}]{footnote1}
\bibinfo{note}{However, one should be aware that the criterion in
  Ref.~\cite{dmrg1} does not apply in most relevant cases, where the rank of
  the reduced density matrix is maximal and consequently the guaranteed upper
  bound would scale exponentially with the system size. Thus, its application
  requires approximations such as~\eqref{approximation}.}
\bibitem[{foo({\natexlab{b}})}]{footnote2}
\bibinfo{note}{This feature survives taking into consideration that we only
  solve the Hamitonian for an odd number of excitations.}
\end{thebibliography}

\end{document}